\documentclass[journal]{IEEEtran}
\usepackage{amssymb}
\usepackage{amsmath}
\usepackage{cite}
\usepackage{url}
\usepackage{xcolor}
\usepackage{cite,graphicx,amsmath,amssymb}
\usepackage{subfigure}
\usepackage{citesort}
\usepackage{fancyhdr}
\usepackage{mdwmath}
\usepackage{mdwtab}
\usepackage{caption}
\usepackage{amsthm}

\newtheorem{remark}{Remark}
\newtheorem{theorem}{Theorem}

\newtheorem{lemma}{Lemma}

\newtheorem{corollary}{Corollary}

\newtheorem{proposition}{Proposition}
\newtheorem{assumption}{Assumption}

\begin{document}
\title{Modeling and Analysis of D2D Millimeter-Wave Networks with Poisson Cluster Processes}

\author{Wenqiang~Yi,~\IEEEmembership{Student Member,~IEEE,}
        Yuanwei~Liu,~\IEEEmembership{Member,~IEEE,}
        and Arumugam~Nallanathan,~\IEEEmembership{Fellow,~IEEE,}

\thanks{This work was presented in part at the IEEE International Conference on Computing, Networking and Communications (ICNC), Maui, Hawaii, USA, March, 2018.}
\thanks{W. Yi, Y. Liu, and A. Nallanathan are with Queen Mary University of London, London, UK (email:\{w.yi, yuanwei.liu, a.nallanathan\}@qmul.ac.uk).}}

\maketitle

\begin{abstract}
  This paper investigates the performance of millimeter wave~(mmWave) communications in clustered device-to-device~(D2D) networks. The locations of D2D transceivers are modeled as a Poisson Cluster Process~(PCP). In each cluster, devices are equipped with multiple antennas, and the active D2D transmitter~(D2D-Tx) utilizes mmWave to serve one of the proximate D2D receivers~(D2D-Rxs). Specifically, we introduce three user association strategies: 1) Uniformly distributed D2D-Tx model; 2) Nearest D2D-Tx model; 3) Closest line-of-site~(LOS) D2D-Tx model. To characterize the performance of the considered scenarios, we derive new analytical expressions for the coverage probability and area spectral efficiency~(ASE). Additionally, in order to efficiently illustrating the general trends of our system, a closed-form lower bound for the special case interfered by intra-cluster LOS links is derived. We provide Monte Carlo simulations to corroborate the theoretical results and show that: 1) The coverage probability is mainly affected by the intra-cluster interference with LOS links; 2) There exists an optimum number of simultaneously active D2D-Txs in each cluster for maximizing ASE; and 3) Closest LOS model outperforms the other two scenarios but at the cost of extra system overhead.
\end{abstract}

\section{Introduction}
 The unparalleled explosion of Internet-enabled mobile devices, applications and services is promoting the development of wireless communication networks. As the spectrum resource is limited in the forth generation (4G) cellular networks, 5G new radio (NR) standard has been considered to be the foundation for the next generation of mobile networks, which works over frequencies both below and above 6 GHz. Apart from extensive researches on sub-6 GHz, such as 2.4 GHz~\cite{chen2016coexistence} and TV white space~\cite{ding2016cellular,peng2011ratio}, millimeter wave (mmWave) has recently received significant attention due to a huge range of free spectrum~\cite{rappaport2014millimeter,pi2011introduction,rappaport2013millimeter}. Numerous protocols show that mmWave frequencies from 30 GHz to 300 GHz have already been utilized in different commercial networks including local area networking in IEEE 802.11ad~\cite{802.11}, personal area networking in IEEE 802.15.3c~\cite{5936164} and fixed-point access links in IEEE 802.16.1~\cite{802.16}. Comparing to traditional networks in 4G, the first distinguishing feature of mmWave networks is the small wave length, which helps to deploy huge antenna arrays at transceivers for enhancing the array gain~\cite{6932503}. This feature reduces inter-cell interferences, the additional noise power and the frequency-dependent path loss~\cite{pi2011introduction}. Another differentiating feature is that mmWave signals are sensitive to blockage effects~\cite{6932503}. Moreover, mmWave signals experience more serious penetration loss than the sub-6 GHz carriers when passing through the blockage~\cite{4460902}. Therefore, the path loss laws for line-of-sight (LOS) links and blockage-dependent non-line-of-sight (NLOS) links are significantly different in mmWave networks~\cite{rappaport2013millimeter,BMP}. A plenty of practical channel measures demonstrate that the path loss exponent of NLOS is more massive than LOS's, because the complicated scattering environment contributes to the severe path loss for NLOS links~\cite{BMP,rappaport201238,deng201528}.

 Accordingly, various articles focus on these two features when modeling mmWave networks. The primary work~\cite{6489099} employed a directional beamforming to fulfill the array gain, but the path loss model was simplified and hence failed to fully reflect mmWave propagation features. Then, site-specific simulation~\cite{330150} and stochastic blockage model~\cite{1296840,6840343,han2017connectivity} were proposed to investigate the performance of mmWave networks with the impact of the blockage. Stochastic geometry is an effective tool to capture the randomness of the networks~\cite{7445146} and recently it was applied in mmWave networks~\cite{6932503,andrews2011tractable}. More particularly, base station locations were modeled as a Poisson Point Process (PPP) on the plane~\cite{andrews2011tractable}. Under this model, a framework combining random blockage process and directional antenna beamforming was designed, which shown a close characterization of the reality~\cite{6932503}.

 However, the aforementioned models only deploy mmWave into a conventional cellular structure where devices download the information from a base station. In this structure, the path loss is serious due to the long distance between transceivers, while mmWave is capable of supporting high rate with short-range networks~\cite{park2007short}. In order to achieve a higher quality cellular network, a key short-distance technology with enormous potential termed device-to-device (D2D) has kindled the interest of academia~\cite{7320989}. To be more specific, D2D networks enable direct links between proximal devices without the aid of cellular networks~\cite{7792168}. When comparing with the traditional architecture in 4G networks, the received power at the intended D2D receiver (D2D-Rx) is typically much higher due to the shorter link distance~\cite{7446343}. With the content centric nature, D2D networks are able to satisfy spatiotemporal correlation in the content demand~\cite{cha2007tube,7797196}. In particular, a user downloads popular files from any of the surrounding transmitters rather than a base station~\cite{6847726,6787081,7150324}. The set of proximate devices is termed a cluster in D2D networks, which corresponds to a hotspot in the heterogeneous cellular networks~\cite{access2010further}.

 The same with mmWave networks, stochastic geometry has also been successfully applied in D2D communications. The primary approach for D2D networks was fixing a D2D transmitter (D2D-Tx) at the origin in a plane and D2D-Rxs were modeled using a PPP~\cite{7320989}. The limitation for this approach is the lack of enough D2D-Txs. As a further development, D2D-Txs were located following a PPP, while D2D-Rxs were modeled as a Poisson Dipole Process (PDP) where every D2D-Tx had a fixed distance to its corresponding D2D-Rx~\cite{6909030,7073589,7147834,7056528}. However, the fixed distance assumption is extremely restrictive. Then the condition was relaxed by assuming that the D2D-Rx was uniformly located within a circle around the serving D2D-Tx~\cite{6928445,6952957,6805658}. Although the distance is variable, the intended D2D-Rx still fails to choose the serving device from multiple proximate transmitters, which is the fundamental nature of D2D networks~\cite{6847726,6787081,7150324}.  Very recently, a realistic tractable D2D structure~\cite{7446343,haenggi2012stochastic} was proposed following a Poisson Cluster Process (PCP)~\footnote{The PCP model is regarded as a promising method for analytically studying various kinds of networks, such as device-to-device, ad hoc network and sensor networks. However, the shortage of experiments in terms of PCP will motivate our future work.}, where the intended user had multiple randomly distributed D2D-Txs and each of them had the ability to be the active serving device. However, this work only focuses on sub-6 GHz networks, while more attention should be paid on the performance of mmWave networks under this architecture as it outperforms sub-6~GHz in short-distance communications.

\subsection{Motivation and Contribution}
As discussed above, mmWave communications have been studied in a variety of scenarios, but there is still short of researches on a short-distance communication system. This shortage motivates us to contribute this treatise. Note that the tractable D2D model mentioned in~\cite{7446343} has a perfect short-distance communication architecture. To increase the capacity and signal-to-interference-plus-noise ratio (SINR) coverage of future wireless networks, it is ideal to deploy mmWave into this D2D structure. Different from~\cite{7446343}, four main issues are carefully addressed in our paper. Firstly, the propagation environment is replaced by two kinds of path loss laws and nakagami-M fading channels due to blockage sensitivity of mmWave signals. Secondly, we employ a sectorial model for analyzing the antenna beamforming. Thirdly, three different user association strategies are proposed to evaluate our system. Lastly, we compare the performance of various carrier frequencies in terms of SINR coverage probability. On the other side, different from PPP modeled mmWave networks~\cite{6932503}, the employment of the PCP results in a unique interference from inter-clusters~\cite{7446343}, which is not negligible in D2D networks. The prime contributions of this paper are as follows:
\begin{itemize}
\item We analyze the coverage performance and area spectral efficiency (ASE) for three different scenarios: \romannumeral1) \emph{Uniform Distribution Model}, where the connected D2D-Tx is uniformly distributed in the same cluster of the typical D2D-Rx; \romannumeral2) \emph{Closest Distribution Model}, where the connected D2D-Tx is the nearest transmitter in the same cluster of the typical D2D-Rx; and \romannumeral3) \emph{Closest LOS Model}, where the connected D2D-Tx is the closest transmitter with an LOS link in the same cluster of the typical D2D-Rx.
\item We characterize the distribution of distances from the typical D2D-Rx to the serving D2D-Tx and intra/inter-interfering devices. Moreover the exact probability density functions (PDFs) of distances for three scenarios are presented.
\item We work out Laplace transforms of intra/inter-cluster interfering powers, using which different coverage probability expressions for three scenarios are derived. Additionally, a closed-form lower bound for an intra-interfered case is presented. We analytically demonstrate that the coverage probability has a positive correlation with the directivity gain at the typical D2D-Rx, while it has the inverse correlation with the number of interfering D2D-Txs. Finally, ASEs are characterized based on the derived coverage probabilities.
\item  We show that: 1) The closest LOS model achieves the best performance among three scenarios regarding the coverage probability; 2) Our model is an interference-limited system due to the content centric nature of D2D communications. In addition, the proposed model is mainly interfered by the intra-devices with LOS links; 3) There is an optimal number of active D2D-Txs in a cluster for achieving the maximum ASE; and 4) Large antenna scale for high frequency has limited impact on SINR coverage in our system. 38 GHz is the best carrier frequency for high SINR regions and 28 GHz is the best for low SINR regions.
\end{itemize}

\subsection{Organization}
The paper is organized as follows. In Section~\ref{System Model}, considering the blockage and antenna beamforming, the clustered device-to-device mmWave communication networks are modeled in a PCP. In Section~\ref{Distance}, we derive distribution expressions of distances from the typical D2D-Rx to the serving D2D-Tx and interfering devices. In Section~\ref{Coverage}, three different distribution scenarios for the serving D2D-Tx are discussed. Coverage probability and ASE algorithms are figured out in this part. In Section~\ref{Numerical}, the numerical results are presented for analyzing and verifying. In Section~\ref{conclusion}, we propose our conclusion.

\section{System Model}\label{System Model}
In this section, we present our system model for appraising the performance of the clustered D2D mmWave communication networks. The paper will focus on downlink coverage probability and ASE. The crucial modeling details are discussed below.

\subsection{Spatial Distribution}
In this treatise, we adopt one of the typical PCP processes, which is a variant of the Thomas cluster process~\cite{haenggi2012stochastic}. More particularly, the devices are located in a group of clusters following a PCP, in which the \emph{parent point process} follows a PPP $\Phi_p$ with a density $\lambda_p$, and the \emph{offspring point processes} with one parent are conditionally independent~\cite{daley2007introduction}. In our system, the centers of clusters $x_c$ contribute to the parent points $x_c\in\Phi_p$, and the devices are offspring points. In each cluster, we assume that all devices, which are independent and identically distributed~(i.i.d.), follow a symmetric normal distribution around the cluster center with mean zero and variance $\sigma^2$. As a result, the location of a device $x_d\in\mathbb{R}^2$ in reference to a cluster center is
\begin{align}
\mathop f\nolimits_{\mathop X\nolimits_d } (\mathop x\nolimits_d ) = \frac{1}{{\mathop {2\pi \sigma }\nolimits^2 }} \exp\left({ { - \frac{{||\mathop x\nolimits_d |{|^2}}}{{\mathop {2\sigma }\nolimits^2 }}} }\right).
\end{align}

\begin{figure*}[t!]
\centering
\subfigure[Graphical illustration of spatial distributions for proposed D2D mmWave networks with the aid of a Poisson Cluster Process.]{\label{PCP} \includegraphics[width= 3.0in, height=2.25in]{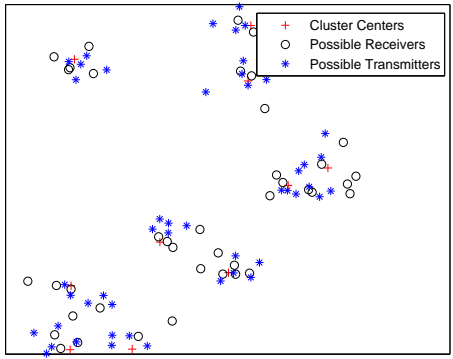}}
\subfigure[Illustration of the stochastic blockage model and beamforming for mmWave networks.]{\label{blockage_model} \includegraphics[width= 3.0in, height=2.25in]{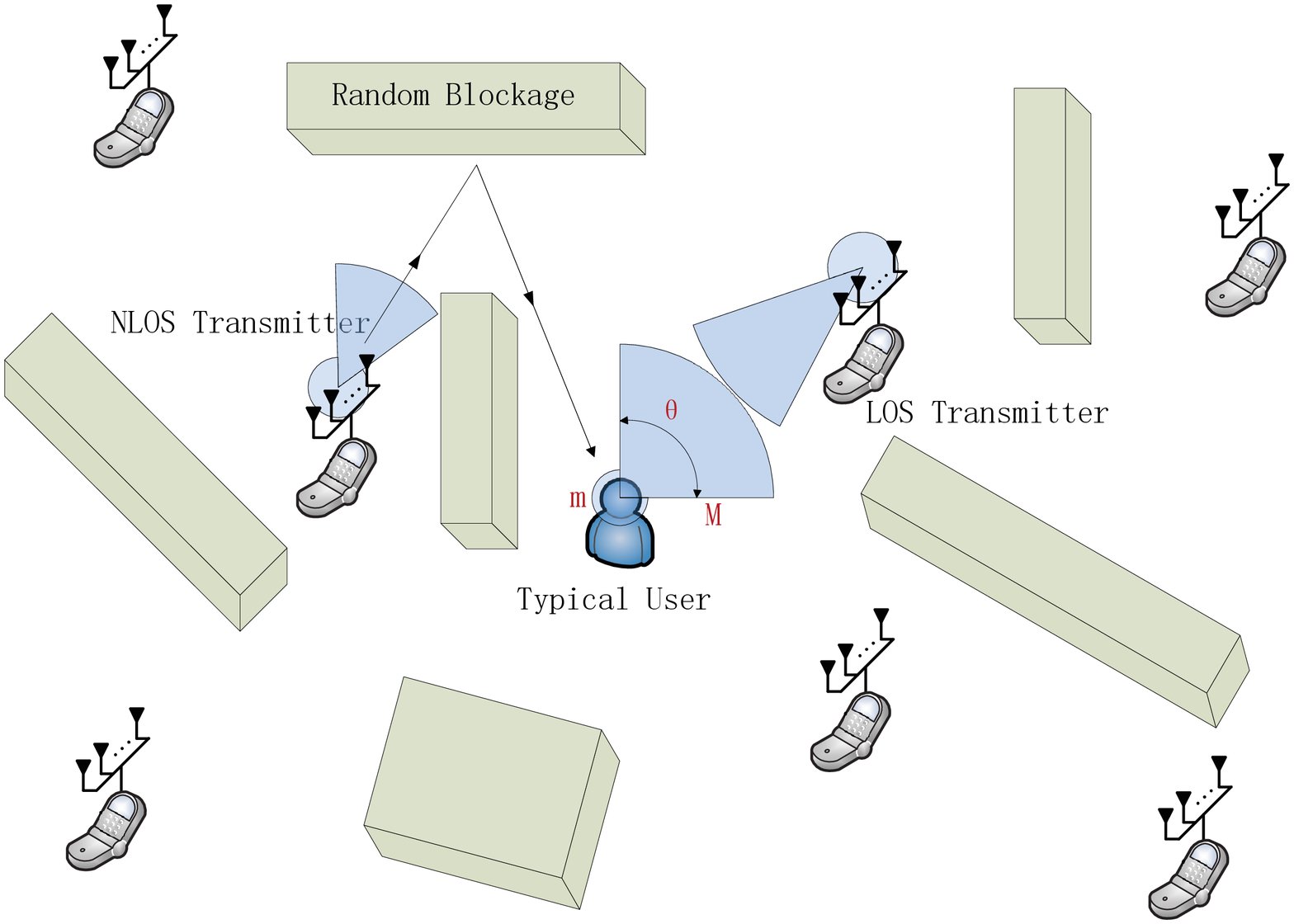}}
\caption{The system model of clustered D2D mmWave networks}
\end{figure*}

For the tractability of analysis, we assumed that the number of devices in every cluster is same with \emph{N} in case one wants to allow all transceivers to communicate simultaneously in a special application~\cite{7446343}. Half of the devices $M=N/2$ are possible transmitters denoted by $\mathbb{N}_t^{{x_c}}$, and the rest $M$ are possible receivers denoted by $\mathbb{N}_r^{{x_c}}$ ($|\mathbb{N}_t^{{x_c}} | = |\mathbb{N}_r^{{x_c}}| = M$). Each transmitter is capable of supporting one receiver at the same time in our model. Although the number of transceivers is fixed, the quantity of simultaneously active transmitters are different across the clusters, which is assumed to be a poisson distribution variable with mean $\bar{s}$ denoted by $\mathop \mathbb{S}\nolimits_t^{\mathop x\nolimits_c }\subseteq\mathop {\rm \mathbb{N}}\nolimits_t^{\mathop x\nolimits_c } $, where $|\mathop \mathbb{S}\nolimits_t^{\mathop x\nolimits_c }|\leq|\mathop {\rm \mathbb{N}}\nolimits_t^{\mathop x\nolimits_c } |$. All these devices are the source of interference except the corresponding transmitter that serving the typical user. Thus, the D2D model is shown in Fig.~\ref{PCP}.

Without loss of generality, we randomly choose one device as a \emph{typical user} that is included in the \emph{typical cluster}. Moreover, the typical user is assumed to be located in the origin of a plane. The center of the \emph{typical cluster} is $\mathop x\nolimits_{c0}  \in {\Phi _p}$ and the transmitters in the typical cluster are denoted by $\mathop \mathbb{N}\nolimits_t^{\mathop x\nolimits_{c0} }$. In the proposed network, the performance of the connection is mainly decided by the distance between the typical user and its \emph{corresponding transmitter}, we provide three different distributions of the corresponding transmitter in the typical cluster for analyzing: 1) \emph{Uniform Distribution Model}: the corresponding transmitter is uniformly distributed in a set of transmitters in the typical cluster; 2) \emph{Closest Distribution Model}: the corresponding transmitter is the closest transmitter in the typical cluster; and 3) \emph{Closest LOS Model}: the corresponding transmitter is the closest transmitter with an LOS link in the typical cluster. Apart from the corresponding transmitter, the rest simultaneously transmitting devices are the source of intra-cluster interference in the typical cluster, so the intra-interfering devices are modeled by a poisson distribution with mean $(\bar{s}- 1)$. However, the active inter-transmitters in every inter-cluster, which contribute to inter-cluster interference, are still poisson distributed with mean $\bar{s}$ as we assumed above.

\subsection{LOS and NLOS Links}
In our system, all transmitters are capable of establishing an LOS or NLOS link to communicate with the typical user when employing mmWave. We assume that the network system is a stochastic blockage model with rectangle Boolean scheme (see Fig.~\ref{blockage_model}), so the probability function of LOS will follow $p(d) = \exp\left( { - \varepsilon d}\right)$, where $\varepsilon$ is determined by the average size and density of blockages, $d$ is the distance between the transmitter and the typical user. In addition, the average LOS distance is $\sqrt{2}/ \varepsilon$~\cite{6840343}. The probability of an LOS link is assumed to be independent with other links. Although LOS probabilities for different links are not independent in reality, ignoring such correlation will cause negligible loss of accuracy in terms of SINR coverage~\cite{6840343} and demonstration will be offered in Section V. Moreover, various path loss $L(d)$ are used to model LOS and NLOS links.
\begin{align}
L(d) = \left\{ {\begin{array}{*{20}{c}}
   {{C_L}{d^{ - {\alpha _L}}},} & {LOS}  \\
   {{C_N}{d^{ - {\alpha _N}}},} & {NLOS}  \\
\end{array}} \right.{\rm{}},
\end{align}
where $\alpha_L$, $\alpha_N$ are LOS and NLOS path loss exponents respectively. $C_L$ is the intercept of LOS links and $C_N$ is that of NLOS links.

\subsection{Directional Beamforming}
We deploy antenna arrays at all transceivers to accomplish directional beamforming as mentioned in~\cite{6932503}. The antenna pattern is assumed to be a sectorial model and the total directivity gain of $l^{th}$ links will be ${G_l} = {G_{{\theta _t},{M_t},{m_t}}}{G_{{\theta _r},{M_r},{m_r}}}$,  where  $G_{{\theta _t},{M_t},{m_t}}$ and $G_{{\theta _r},{M_r},{m_r}}$ are antenna gains at transmitters and receivers, respectively. In each antenna, $\theta_s$ ($s\in\{t,r\}$) is the main lobe beamwidth, then $M_s$ and $m_s$  denote the directivity gain of main lobe and back lobe. Note that angles of arrival and angles of departure for all LOS and NLOS links are independently and uniformly distributed in the range $[0,2\pi]$, so random directivity gains ${G_l}$ of interferences have $N_G=4$ patterns with the value $a_i$ and probability $b_i$, where $i \in \left\{ {1,2,3,4} \right\}$. See Table.~\ref{tab1}.

\begin{table}[h]
\centering
\caption{Probability and Value of $G_l$}
\label{tab1}
\begin{tabular}{|c|c|c|c|c|}
\hline
i     &1        &2        &3        &4\\ \hline
$a_i$ &$M_tM_r$ &$m_tM_r$ &$M_tm_r$ &$m_tm_r$\\  \hline
$b_i$ &$\frac{{{\theta _t}}}{{2\pi }}\frac{{{\theta _r}}}{{2\pi }}$ &$(1 - \frac{{{\theta _t}}}{{2\pi }})\frac{{{\theta _r}}}{{2\pi }}$ &$\frac{{{\theta _t}}}{{2\pi }}(1 - \frac{{{\theta _r}}}{{2\pi }})$ &$(1 - \frac{{{\theta _t}}}{{2\pi }})(1 - \frac{{{\theta _r}}}{{2\pi }})$\\ \hline
\end{tabular}
\end{table}

For different carrier frequencies, the antenna array should be changeable since higher frequencies allow manufacturing more antenna elements for compensating the possible higher path-loss. Under this condition, when analyzing various carrier frequencies, we will change $M_s$ and $m_s$ into $N_aM_s$ and $N_am_s$, respectively, where $N_a$ respects the number of antennas assembled at devices.

\subsection{Channel Model}
Assuming the corresponding transmitter is located at $x_{d0}$ to the center of the typical cluster, the distance between the typical user and the corresponding transmitter is $||x_{c0}+x_{d0}||$ ($x_{c0}\in\Phi_p$, $x_{d0}\in\mathop \mathbb{N}\nolimits_t^{\mathop x\nolimits_{c0} }$). The received power of the typical link is given by
\begin{align}
{{\rm{P}}_r}{\rm{ = }}{G_0{\rm{P}}_0}|{h_l}{|^2}L(||x_{c0}+x_{d0}||),
\end{align}
where $P_0$ is the transmitting power of each device, $h_l$ is the small fading term for $l^{th}$ link and $h_l\sim$ independent Nakagami fading. As a result, $|h_l|^2$ follows a normalized Gamma random variable. The Nakagami fading parameters are $N_L$ and $N_N$ for LOS and NLOS links, respectively. $N_L$ and $N_N$ are assumed to be positive integers for simplicity~\cite{6932503}.

In our model, the interferences have two sources. One is intra-cluster interference $I_{intra}$ from the typical cluster, and the other is inter-cluster interference $I_{inter}$ from other clusters. The distance between the typical user and the transmitter in the typical cluster is $||x_{c0}+x_{d}||$ ($x_{c0}\in\Phi_p$, $x_d\in\mathop \mathbb{N}\nolimits_t^{\mathop x\nolimits_{c0} }$), and that from the typical user to the transmitter in other clusters is $||x_{c}+x_{d}||$ ($x_{c}\in\Phi_p$, $x_d\in\mathop \mathbb{N}\nolimits_t^{\mathop x\nolimits_{c} }$). The two kinds of different interference power are expressed as follows
\begin{align}
{I_{intra}} = \sum\limits_{\mathop x\nolimits_d  \in \mathop \mathbb{S}\nolimits_t^{\mathop x\nolimits_{c0} } \backslash \mathop x\nolimits_{d0} } {{G_l}{P_0}} |{h_l}{|^2}L({||x_{c0}+x_{d}||}),
\end{align}
\begin{align}
{I_{inter}} = \sum\limits_{\mathop x\nolimits_c  \in \mathop \Phi \nolimits_p^{} \backslash \mathop x\nolimits_{c0} } {\sum\limits_{\mathop x\nolimits_d  \in \mathop \mathbb{S}\nolimits_t^{\mathop x\nolimits_c } } {{G_lP_0}|{h_l}{|^2}L({||x_{c}+x_{d}||})} }.
\end{align}
As a result, the SINR at the typical user is given by
\begin{align}
SINR = \frac{{{P_r}}}{{\sigma _n^2 + {I_{intra}} + {I_{inter}}}},
\end{align}
where $\sigma _n^2$ is the thermal noise power normalized by $P_0$. The power of transmitter $P_0$ can be canceled in SINR. Without any loss of generality, we assume $P_0=1$.

\section{Distribution of Distances}\label{Distance}
 We will discuss the distribution of the distances between the typical user and other devices in this section. Before that, we introduce two different distributions as mentioned in~\cite{7446343} below in order to simplify the notation.

\emph{Rayleigh Distribution}: the probability density function (PDF) is defined as $Ra(x,{\sigma ^2})$
\begin{align}
Ra(x,{\sigma ^2}) = \frac{x}{{{\sigma ^2}}} \exp \left({ { - \frac{{{x^2}}}{{2{\sigma ^2}}}} }\right),x > 0,
\end{align}
where $\sigma$ is the scale parameter of Rayleigh distribution.

\emph{Rician Distribution}: the PDF is defined as $Ri(x,y,{\sigma ^2})$
\begin{align}
Ri(x,y,{\sigma ^2}) = \frac{x}{{{\sigma ^2}}} \exp \left({ { - \frac{{{x^2} + {y^2}}}{{2{\sigma ^2}}}}}\right) {I_0}\left( {\frac{{xy}}{{{\sigma ^2}}}} \right),x > 0,
\end{align}
where $\sigma$ is the scale parameter of Rician distribution and $I_0(.)$ is the first kind Modified Bessel Function with zero order.

\subsection{Distribution in Uniform Distribution Model}
In this part, the distribution of distances in uniform distribution model will be characterized. We will start the demonstration with the typical cluster and then other clusters.
\subsubsection{Distance Distribution in Typical Cluster}
Assuming the set of distances between the typical user and the possible transmitters in the typical cluster is ${\left\{ {{D_i}} \right\}_{i = 1:M}}$ denoted by $\mathbb{D} _t^{{x_{c0}}}$ (${D_i}\in\mathbb{D} _t^{{x_{c0}}}$). $d_i$ is the realization of $D_i$ and ${d_i} = ||{x_{c0}} + {x_d}||$ ($x_{c0}\in\Phi_p$, $x_d\in\mathop \mathbb{N}\nolimits_t^{\mathop x\nolimits_{c0} }$). Since $x_{c0}$ and $x_d$ are Gaussian Random Variables (i.i.d.) with $\sigma ^2$ variance, $d=({x_{c0}} + {x_d})$ is a Gaussian Random Variable with $2\sigma ^2$ variance so that $D_i$ can be approximated by a PDF of ${f_D}(d) = Ra(d,2{\sigma ^2})$. However, $d=||{x_{c0}} + {x_d}||$ is conditional on the distance ${v_{c0}} = ||{x_{c0}}||$ because the transceivers are i.i.d around the cluster center in our system model. Therefore the exact PDF is shown as below~\cite[Lemma 1]{7446343}:
\begin{align}\label{distribution}
{f_D}(d|{v_{c0}}) = Ri(d,{v_{c0}},{\sigma ^2}).
\end{align}

In typical cluster, since $M$ elements of $\mathbb{D} _t^{{x_{c0}}}$ are i.i.d and the corresponding transmitter is selected uniformly at random, all distributions of distances including the corresponding transmitter and intra-interfering devices will follow Rician distribution in \eqref{distribution}. The results are shown formally as below.

 \emph{The distance of the typical link}: the distance between the typical user and its corresponding transmitter is assumed to be $r_0=||x_{c0}+x_{d0}||$ ($x_{d0}\in\mathop \mathbb{N}\nolimits_t^{\mathop x\nolimits_{c0} }$). As mentioned above, the PDF of typical link distance is $f_R(r_0|v_{c0})=Ri(r_0,v_{c0},\sigma^2)$.

\emph{Distances of intra-cluster interfering links}: The distance from intra-cluster interfering device to the typical user is $s_a=||x_{c0}+x_{d}||$ (${x_d} \in \mathbb{S}_t^{{x_{c0}}}\backslash {x_{d0}}$). Utilizing the same method discussed in \eqref{distribution}, the PDF of this case is $f_S(s_a|v_0)=Ri(s_a,v_0,\sigma^2)$.
\subsubsection{Distance Distribution in Other Clusters}
In other clusters, the set of distances between the typical user and the possible transmitter is ${\left\{ {{U_i}} \right\}_{i = 1:M}}$ denoted by $\mathbb{U} _t^{{x_{c}}}$ (${U_i}\in\mathbb{U} _t^{{x_{c}}}$). $u_i$ is the realization of $U_i$ and ${u_i} = ||{x_{c}} + {x_d}||$ ($x_{c}\in\Phi_p$, $x_d\in\mathop \mathbb{N}\nolimits_t^{\mathop x\nolimits_{c} }$). $x_d$ has the same distribution with that in the typical cluster, and the only difference is that ${u} = ||{x_{c}} + {x_d}||$ is conditional on the distance $v_c$ ($v_c=||x_c||$). Evidently, the PDF of distances from the typical user to simultaneous transmitters in other clusters is as follows~\cite[Lemma 2]{7446343}
\begin{align}\label{distribution2}
{f_U}(u|{v_{c}}) = Ri(u,{v_{c}},{\sigma ^2}).
\end{align}

The set of distances of inter-cluster interfering links between the typical user and simultaneously transmitting devices in other clusters is denoted by $w_i=||x_{c}+x_{d}||$ ($x_{d}\in\mathop \mathbb{S}\nolimits_t^{\mathop x\nolimits_{c} }$). It is conditioned on the distance ${v_{c}} = ||{x_{c}}||$. As the inter-interfering device is selected at random, $w_i$ has the same distribution with $u_i$. The distances $w_a=||x_{c}+x_{d}||$ of inter-cluster interfering link will follow $f_W(w_a|v_c)=Ri(w_a,v_c,\sigma^2)$.

\begin{remark}
As the corresponding transmitter is located in the typical cluster, there is no difference among three scenarios in terms of distances distribution in inter-clusters. Therefore, the distance of inter-cluster interfering links in other two scenarios are same with uniform distribution model, and we will omit this in the following discussion.
\end{remark}

\subsection{Distribution in Closest Distribution Model}
In this part, the distribution of distances in closest distribution model will be discussed. Unlike the uniform distribution model above, we assume that the corresponding transmitter in the typical cluster is the nearest one with $r_1=||x_{c0}+x_{d1}||$ ($x_{c0}\in\Phi_p$, ${x_{d1}} = \min \left( {\mathbb{S}_t^{{x_{c0}}}} \right)$ ). In this model, the distribution of distance for the closest link is shown below.

\emph{The distance of the closest typical link}: The distance from the nearest transmitter to the typical user $r_1$ is conditioned on the distance ${v_{c0}} = ||{x_{c0}}||$ and the PDF is easy to be deduced from \cite[Lemma 3]{7446343}. We choose the $1$st-closest content available strategy here and the equation is shown below
\begin{align}\label{closest}
{f_{{R_1}}}({r_1}|{v_{c0}}) = MQ_m{\left( {\frac{{{v_{c0}}}}{\sigma },\frac{{{r_1}}}{\sigma }} \right)^{M - 1}}{f_{R_1}}\left( {{r_1}|{v_{c0}}} \right),
\end{align}
\emph{where $Q_m\left( {x,y} \right) = \int_y^\infty  {t \exp\left({ { - \frac{{{t^2} + {x^2}}}{2}} }\right)} {I_0}\left( {xt} \right)dt$, and ${f_{{R_1}}}({r_1}|{v_{c0}})=Ri\left( {{r_1},{v_{c0}},{\sigma ^2}} \right)$.}

As the corresponding transmitter is the closest one, the rest distances $s_{r}=||x_{c0}+x_{d}||$ (${x_d} \in \mathbb{S}_t^{{x_{c0}}}\backslash {x_{d1}}$) from the typical user to intra-interfering transmitters are larger than $r_1$. They are conditioned on the distance $r_1$ and $v_0$. The PDF is illustrated below.

\emph{Distances of intra-interfering links}: The set of distances between the typical user and the rest intra-interfering devices $s_r$ is conditioned on the distance ${v_{c0}} = ||{x_{c0}}||$ and the closest distance $r_1$, it is shown below \cite[Lemma 4]{7446343}
\begin{align}\label{nearest}
{f_{{S_{r}}}}({s_{r}}|{v_{c0}},{r_1}) = \left\{ {\begin{array}{*{20}{c}}
   {\frac{{{f_{{S_{r}}}}({s_{r}}|{v_{c0}})}}{{Q_m\left( {\frac{{{v_{c0}}}}{\sigma },\frac{{{r_1}}}{\sigma }} \right)}},} & {{s_{r}} > {r_1}}  \\
   0 & {{s_{r}} \le {r_1}}  \\
\end{array}} \right.,
\end{align}
\emph{where ${f_{{S_r}}}({s_r}|{v_{c0}}) = Ri\left( {{s_r},{v_{c0}},{\sigma ^2}} \right)$.}

\subsection{Distribution in Closest LOS Model}
Different with closest distribution model, we focus on the nearest device with an LOS link in closest LOS model. The set of transmitters with LOS links in the typical cluster is denoted by $\mathbb{S}_{{t_L}}^{{x_{c0}}}\subset{\mathbb{S}_t^{{x_{c0}}}}$. On the other hand, the set of NLOS is $\mathbb{S}_{{t_N}}^{{x_{c0}}}\subset{\mathbb{S}_t^{{x_{c0}}}}$. And the corresponding transmitter is the nearest one with an LOS link, which has the distance $r_L=||x_{c0}+x_{d_L}||$ ($x_{c0}\in\Phi_p$, ${x_{d_L}} = \min \left( \mathbb{S}_{{t_L}}^{{x_{c0}}} \right)$ ). Note that the probability function of LOS will follow $p(d) = \exp\left({ - \varepsilon d}\right)$, the distribution of distance $r$ between the transmitter with an LOS link and the typical user in the typical cluster is
\begin{align}\label{closestLOS}
{F_L}\left( {r|{v_{c0}}} \right) = \exp \left( { - \varepsilon r} \right){f_D}\left( {r|{v_{c0}}} \right).
\end{align}

Under this condition, the distance for the closest LOS link is distributed as below.
\begin{lemma}
\emph{(The distance of the closest typical LOS link): The distance of the nearest transmitter with an LOS link $r_L$ is conditioned on the distance ${v_{c0}} = ||{x_{c0}}||$ and the PDF is}
\begin{align}
{f_{{R_L}}}({r_L}|{v_{c0}}) = M{\left( {1 - \int_0^{{r_L}} {{F_L}\left( {r|{v_{c0}}} \right)dr} } \right)^{M - 1}}{F_L}\left( {{r_L}|{v_{c0}}} \right).
\end{align}
\begin{IEEEproof}
We randomly choose one device in the typical cluster to be the corresponding transmitter. It has a distance $r_L$ to the typical user. Note that the typical cluster has $M$ transmitters, so there are $(M-1)$ transmitters located beyond the circle with the radius of $r_L$. With the aid of \eqref{closestLOS}, the distance distribution of the closest typical LOS link is derived as above.
\end{IEEEproof}
\end{lemma}

As we discussed above, the rest distances with LOS links $s_{L}=||x_{c0}+x_{d}||$ (${x_d} \in \mathbb{S}_{{t_L}}^{{x_{c0}}}\backslash {x_{d_L}}$) from the typical user to simultaneously transmitting devices in the typical cluster must be larger than $r_L$. They are conditioned on the typical link distance $r_L$ and the distance $v_0$. The distribution of rest distances is expressed as below.
\begin{lemma}
\emph{(Distances of intra-interfering LOS links): The distance of the rest typical LOS links $s_L$ are conditioned on the distance ${v_{c0}} = ||{x_{c0}}||$ and the closest distance with an LOS link $r_L$, it is}
\begin{align}\label{nearestLOS}
{f_{{S_{L}}}}({s_{L}}|{v_{c0}},{r_L}) = \left\{ {\begin{array}{*{20}{c}}
   {\frac{{{f_{{R}}}({s_{L}}|{v_{c0}})}}{{Q_m\left( {\frac{{{v_{c0}}}}{\sigma },\frac{{{r_L}}}{\sigma }} \right)}},} & {{s_{L}} > {r_L}}  \\
   0 & {{s_{L}} \le {r_L}}  \\
\end{array}} \right..
\end{align}
\begin{IEEEproof}
 As the locations of LOS transmitters, which follows Rician Distribution, are independent of NLOS devices, \eqref{nearest} also exists in this case. Obviously, if the distance $s_r$ is less than $r_L$, the probability should be 0.
\end{IEEEproof}
\end{lemma}

Different with LOS links, the distance of devices in the typical cluster with NLOS links are distributed as that in uniform distribution model.

\section{Coverage Probability and Area Spatial Efficiency Analysis}\label{Coverage}
In this section, we focus on the coverage probability and ASE in different scenarios depending on the distances distribution.

\subsection{Uniform Distribution Model}
In this model, like various strategies have been proposed in recent articles, for example, \emph{Uniform Content Availability} in D2D networks~\cite{7446343} and \emph{RNRF Selection Scheme} in NOMA networks~\cite{7445146}, the typical user will choose the corresponding transmitter randomly in the typical cluster. This strategy offers a fair opportunity for each device to access the content in the cluster. The benefit of this user association scheme is that networks do not need the additional knowledge of instantaneous channel state information (CSI) which is not available on some networks due to the poor performance of the basic equipment. To make tractable calculation, we first introduce \emph{Laplace Transform} to figure out the expected value of interference. Then the coverage probability will be derived using \emph{Laplace Transform of Interference}.
\subsubsection{Laplace Transform of Interference}
We first derive analytical expressions and approximations on the Laplace transform of intra-cluster interference. As in the real-life world, the number of active D2D pairs is far less than the number of possible transceivers in most clusters. For example, assuming that people in a library form a cluster, the scale of devices sharing study materials simultaneously is much smaller compared with the number of customers because the majority of tasks in a library should be reading and self-study rather than transmitting information. Therefore, we add the assumption $M\gg\bar{s}$ in the following content for the tractability of analysis.

\begin{lemma}\label{lemma3}
\emph{(Laplace transform of interference in the typical cluster): When $M\gg\bar{s}$, the $n$th conditional Laplace transform of interference in the typical cluster is}
\begin{align}\label{Lin}
\begin{split}
L_{{I_{intra}}}^n(s|{v_{c0}}) = &{\exp\bigg({ - (\bar s - 1)}}\\
&\times{{\int_0^\infty  {\left( {Q(s,r) + Z(s,r)} \right)} }}{{ {f_R}(r|{v_{c0}})dr}\bigg)},
\end{split}
\end{align}
where
\begin{align}\label{Qfunction}
Q(s,r) = \left( {1 - \sum\limits_{i = 1}^{N_G} {\frac{{{b_i}{N_L}^{{N_L}}}}{{{{({a_i}ns{C_L}{r^{ - {\alpha _L}}} + {N_L})}^{{N_L}}}}}} } \right){\exp\left({ - \varepsilon r}\right)},
\end{align}
\begin{align}\label{Zfunction}
\begin{split}
Z(s,r) = &\left( {1 - \sum\limits_{i = 1}^{N_G} {\frac{{{b_i}{N_N}^{{N_N}}}}{{{{({a_i}ns{C_N}{r^{ - {\alpha _N}}} + {N_N})}^{{N_N}}}}}} } \right)\\
&\times(1 - {\exp\left({ - \varepsilon r}\right)}).
\end{split}
\end{align}
\begin{IEEEproof}
See Appendix~A.
\end{IEEEproof}
\end{lemma}
\begin{remark}\label{remark_2}
Conditioning on the certain $s$ and $r$, $Q(s,r)$ and $Z(s,r)$ represent the probabilities of intra-cluster interferences from LOS and NLOS links, respectively. Additionally, it is easy to infer that (17) is a monotonic decreasing function with $r$, which means in short-range clustered networks the probability of intra-cluster LOS interference will be high.
\end{remark}

Note that the assumption $M\gg\bar{s}$ is applicable when the number of simultaneously transmitting devices per cluster is much smaller than the cluster size. As analyzed in the sequel, the assumption is also the regime where the networks will be optimized in terms of ASE. Therefore, the simpler expression will be used as a proxy of exact expression for the analytical equations and approximations.

\begin{assumption}\label{assumption1}
 \emph{Although distribution of distance from intra-transmitters to the typical user is conditioned on the distance $v_{c0}$, the analysis is essentially simplified by ignoring this condition. We assume the distance between the typical user and the corresponding transmitter follows ${f_R}(r) = Ra(r,2{\sigma ^2})$ as we discussed in the distribution of uniform distribution model.}
\end{assumption}

This assumption is under the consideration that regarding the distribution of intra-cluster devices, the conditioning on the distance $v_{c0}$ is weak enough to be ignored~\cite{7446343}. It is treated as a tight approximation for the following calculation so that more insights can be directly obtained from the analytical results.

\begin{corollary}
\emph{(Approximation): Under the \textbf{Assumption~\ref{assumption1}}, the Laplace transform of interference in the typical cluster is approximated as below}
\begin{align}\label{Lina}
\begin{split}
\tilde{L}_{I_{\mathop{intra}}}^n(s)=&\exp\bigg(-(\bar s - 1)\\
 &\times{{ { \int_0^\infty  {\left( {Q(s,r) + Z(s,r)} \right)} {f_R}(r)dr} }\bigg)}.
\end{split}
\end{align}
\end{corollary}

After the analysis of interference in the typical cluster, we now state the analytical result for the Laplace transform of inter-cluster interference.
\begin{lemma}\label{lemma4}
\emph{(Laplace transform of interference in other clusters): When $M\gg\bar{s}$, the $n$th conditional Laplace transform of interference in inter-clusters is}
\begin{align}\label{Lout}
\begin{split}
&L_{I_{\mathop{inter}}}^n(s)=\exp\bigg({ - 2\pi {\lambda _p}\int_0^\infty  {\bigg( {1 -  {\exp\bigg({ { - \bar s}}}}}}\\
&\times {{{{{{\int_0^\infty  {\left( {Q(s,w) + Z(s,w)} \right){f_W}\left( {w|v} \right)dw} } }\bigg)} } \bigg)} vdv}\bigg).
\end{split}
\end{align}
\begin{IEEEproof}
See Appendix~B.
\end{IEEEproof}
\end{lemma}
\begin{remark}\label{remark_inter}
Note that the inter-interfering devices are distributed uniformly at random, which means that the above expression is applicable for three proposed scenarios.
\end{remark}
\subsubsection{Coverage Probability}
We set a pre-decided threshold of SINR $\gamma_{th}$ to analyze the performance. The SINR that exceeds $\gamma_{th}$ contributes to the coverage probability. It is expressed as
\begin{align}\label{P}
\begin{split}
P_u=&{P_L}\left\{ {\frac{{{G_0}|{h_0}{|^2}{C_L}{r_0}^{ - {\alpha _L}}}}{{\left( {\sigma _n^2 + {I_{intra}} + {I_{inter}}} \right)}} > \gamma_{th} } \right\}\\
&+ {P_N}\left\{ {\frac{{{G_0}|{h_0}{|^2}{C_N}{r_0}^{ - {\alpha _N}}}}{{\left( {\sigma _n^2 + {I_{intra}} + {I_{inter}}} \right)}} > \gamma_{th} } \right\},
\end{split}
\end{align}
where $r_0=||x_{c0}+x_{d0}||$. $P_L\{.\}$ and $P_N\{.\}$ are probabilities of the typical LOS link and NLOS link respectively. With the aid of Laplace transform, the tight upper bound expression for $P_u$ is shown in the \textbf{Theorem~\ref{theorem1}}.

\begin{theorem}\label{theorem1}
\emph{Using the Laplace transform of intra-cluster interference \textbf{Lemma~\ref{lemma3}} and inter-cluster interference \textbf{Lemma~\ref{lemma4}}, we figure out the tight upper bound for the coverage probability under uniform distribution model. It is given by}
\begin{align}\label{uniPc}
\begin{split}
P_u < &\int\limits_0^\infty  {\int\limits_0^\infty  {\left( {X({r_0},{v_{c0}}) + Y({r_0},{v_{c0}})} \right)} } \\
&\times {f_R}({r_0}|{v_{c0}}){f_{{V_{c0}}}}({v_{c0}})d{r_0}d{v_{c0}},
\end{split}
\end{align}
\emph{where}
\begin{align}
\begin{split}
&X({r_0},{v_{c0}}) =\\
& \sum\limits_{n = 1}^{{N_L}} {{{( - 1)}^{n + 1}}{ N_L \choose n}}\exp\left({-\varepsilon r_0}\right) \exp\left({ { - \frac{{\gamma_{th} {r_0}^{{\alpha _L}}{\eta _L}}}{{{C_L}{G_0}}}n\sigma _n^2} }\right)\\
&\times L_{{I_{intra}}}^n\left( {\frac{{\gamma_{th} r_0^{{\alpha _L}}{\eta _L}}}{{{C_L}{G_0}}}|{v_{c0}}} \right)L_{{I_{inter}}}^n\left( {\frac{{\gamma_{th} r_0^{{\alpha _L}}{\eta _L}}}{{{C_L}{G_0}}}} \right),
\end{split}
\end{align}
\begin{align}\label{Y}
\begin{split}
&Y({r_0},{v_{c0}}) = \sum\limits_{n = 1}^{{N_N}}{{( - 1)}^{n + 1}}  \\
&\times {{ N_N \choose n}}(1-\exp\left({-\varepsilon r_0}\right)) \exp\left( { { - \frac{{\gamma_{th} {r_0}^{{\alpha _N}}{\eta _N}}}{{{C_N}{G_0}}}n\sigma _n^2}}\right)\\
 &\times L_{{I_{intra}}}^n\left( {\frac{{\gamma_{th} r_0^{{\alpha _N}}{\eta _N}}}{{{C_N}{G_0}}}|{v_{c0}}} \right)L_{{I_{inter}}}^n\left( {\frac{{\gamma_{th} r_0^{{\alpha _N}}{\eta _N}}}{{{C_N}{G_0}}}} \right),
 \end{split}
\end{align}
\emph{and $f_{{V_{c0}}}({v_{c0}})=Ra(v_{c0},\sigma ^2)$. $\eta_L  = N_L{\left( {N_L!} \right)^{ - \frac{1}{N_L}}}$, $\eta_N  = N_N{\left( {N_N!} \right)^{ - \frac{1}{N_N}}}$.}
\begin{IEEEproof}
See Appendix~C.
\end{IEEEproof}
\end{theorem}

\begin{remark}
The results derived in \textbf{Theorem~\ref{theorem1}} show that the coverage probability of uniform distribution model $P_u$ is independent of the cluster size $M$.
\end{remark}

\begin{corollary}\label{appro}
\emph{(Approximation of coverage probability): Under \textbf{Assumption~\ref{assumption1}} the tight upper bound for the probability of coverage in uniform distribution model is}
\begin{align}
P_u^a<{\int\limits_0^\infty  {\left( {W({r_0}) + K({r_0})} \right)} } {f_{{R}}}(r_0)d{r_0},
\end{align}
\emph{where}
\begin{align}
\begin{split}
&W({r_0}) =\\
&\sum\limits_{n = 1}^{{N_L}} {{{( - 1)}^{n + 1}}{ N_L \choose n }} \exp\left({-\varepsilon r_0}\right) \exp\left({ { - \frac{{\gamma_{th} {r_0}^{{\alpha _L}}{\eta _L}}}{{{C_L}{G_0}}}n\sigma _n^2} }\right)\\
&\times \tilde L_{{I_{intra}}}^n\left(\frac{{\gamma_{th} r_0^{{\alpha _L}}\eta_L}}{{{C_L G_0}}}\right)L_{{I_{inter}}}^n\left(\frac{{\gamma_{th} r_0^{{\alpha _L}}\eta_L}}{{{C_L G_0}}}\right),
\end{split}
\end{align}
\begin{align}
\begin{split}
&K({r_0}) = \sum\limits_{n = 1}^{{N_N}}{{( - 1)}^{n + 1}}\\
&\times {{ N_N \choose n }} (1-\exp\left({-\varepsilon r_0}\right)) \exp\left({ { - \frac{{\gamma_{th} {r_0}^{{\alpha _N}}{\eta _N}}}{{{C_N}{G_0}}}n\sigma _n^2} }\right)\\
 &\times \tilde L_{{I_{intra}}}^n\left(\frac{{\gamma_{th} r_0^{{\alpha _N}}\eta_N}}{{{C_N G_0}}}\right)L_{{I_{inter}}}^n\left(\frac{{\gamma_{th} r_0^{{\alpha _N}}\eta_N}}{{{C_N G_0}}}\right).
\end{split}
\end{align}
\begin{IEEEproof}
The proof is same with \textbf{Theorem~\ref{theorem1}}, except that it is no conditioned with distance $v_{c0}$. As a result, we use \eqref{Lina} instead of \eqref{Lin}.
\end{IEEEproof}
\end{corollary}

\begin{assumption}\label{assumption2}
\emph{We assume a special case with small $\sigma$ and large $\bar{s}$. As discussed in~\cite{7446343}, the coverage probability of the typical user is mainly affected by the intra-cluster interference under such condition, and thus the received interfering signals from inter-transmitters are ignored. Additionally, NLOS signals and noise are negligible in our system due to the nature of mmWave networks~\cite{6932503} and content centric property of D2D architecture~\cite{7446343}, respectively. In a word, under this assumption, we only consider the intra-cluster interference and all links are regarded as LOS.}
\end{assumption}

This assumption is a common scene in our real world. For example, people watching the football game sitting around a screen in a bar constitute a cluster. They are able to use the mobile devices to share the short video of the game's highlights by D2D with mmWave. In this case, the distance between the persons in this cluster is short, which means $\sigma$ is small. Additionally, there is a plenty of devices sharing content simultaneously during the peak-time, so $\bar{s}$ is large. Therefore, \textbf{Assumption~2} is a reasonable simplification for analysis our system.

In this assumption, we have ignored the inter-cluster interference so that it seems like only a certain cluster is taken into account, which is no different from traditional D2D networks. However, in fact, the typical user is randomly chosen across all clusters in our system, which means each cluster has the fair probability to be the typical cluster. This feature ensures that the analysis under \textbf{Assumption~2} is still able to reflect the whole picture of clustered D2D mmWave networks that have the significantly different clustered structure with traditional D2D communications. Moreover, since we utilize mmWave as our carrier frequency, the propagation environment and antenna pattern are totally different from the traditional ones as well.

With the condition of \textbf{Assumption~\ref{assumption2}}, the Laplace transform of interference in the typical cluster is shown as follows.
\begin{lemma}
\emph{(Laplace transform of interference in the typical cluster): When $M\gg\bar{s}$, the $n$th conditional Laplace transform of interference in the typical cluster under \textbf{Assumption~\ref{assumption2}} is}
\begin{align}\label{LinLa}
\ddot{L}_{I_{\mathop{intra}}}^n(s|{v_{c0}})= \exp\left({  - ( {\bar s - 1} )\int_{{0}}^\infty  {Q_a\left(s, {{r}} \right){f_{{R}}}({r}|{v_{c0}})d{r}}}\right),
\end{align}
where $Q_a(s,r) = \left( {1 - \sum\limits_{i = 1}^{N_G} {\frac{{{b_i}{N_L}^{{N_L}}}}{{{{({a_i}ns{C_L}{r^{ - {\alpha _L}}} + {N_L})}^{{N_L}}}}}} } \right)$.
\begin{IEEEproof}
Under the \textbf{Assumption~\ref{assumption2}}, there is no noise and only LOS interference in the typical cluster. We should remove all the useless parts from \eqref{Lin}.
\end{IEEEproof}
\end{lemma}

\begin{corollary}
\emph{(Approximation): Combining with the \textbf{Assumption~\ref{assumption1}}, the approximate Laplace transform of interference in the typical cluster is}
\begin{align}\label{Linau}
\hat{L}_{I_{\mathop{intra}}}^n(s)= {\exp\left({ { - (\bar s - 1)\int_0^\infty  { Q_a(s,r) } } {f_R}(r)dr} \right)}.
\end{align}
\begin{IEEEproof}
The proof procedure is similar as \eqref{Lina} and hence is skipped here.
\end{IEEEproof}
\end{corollary}

\begin{theorem}
\emph{With the aid of Laplace transform of intra/inter interference discussed above, we derive the tight upper bound for the coverage probability for uniform distribution model under \textbf{Assumption~\ref{assumption2}}. It follows}
\begin{align}\label{LOSPc1}
P_s<\int\limits_0^\infty  {\int\limits_0^\infty  {{\ddot{X}}\left( {{r_0},{v_{c0}}} \right)} } {f_{{R}}}({r_0}|{v_{c0}}){f_{{V_{c0}}}}({v_{c0}})d{r_0}d{v_{c0}},
\end{align}
\emph{where ${\ddot{X}}\left( {{r_0},{v_{c0}}} \right)= \sum\limits_{n = 1}^{{N_L}} {{{( - 1)}^{n + 1}}{ N_L \choose n }} \ddot{L}_{{I_{{\rm{int}}ra}}}^n\left(\frac{{\gamma_{th} r_0^{{\alpha _L}}\eta_L}}{{{C_L G_0}}}|{v_{c0}}\right). $}
\begin{IEEEproof}
Same with \textbf{Theorem~\ref{theorem1}}, but there is only LOS intra-cluster interference existed.
\end{IEEEproof}
\end{theorem}

The different $N_a$ for various carrier frequencies can be canceled from (30), which means under this assumption, the SINR coverage probability for various carrier frequencies has no relationship with the scale of antenna arrays.

\begin{corollary}
\emph{(Approximation of coverage probability): under \textbf{Assumption~\ref{assumption1}} and \textbf{Assumption~\ref{assumption2}} the tight upper bound for the probability of coverage is shown as below.}
\begin{align}
P_s^a<\int\limits_0^\infty  {\ddot{W}({r_0})} f_{{R}}({r_0})d{r_0},
\end{align}
\emph{where $\ddot{W}({r_0}) =\sum\limits_{n = 1}^{{N_L}} {{{( - 1)}^{n + 1}}{ N_L \choose n }} \hat{L}_{{I_{intra}}}^n\left(\frac{{\gamma_{th} r_0^{{\alpha _L}}\eta_L}}{{{C_L G_0}}}\right).$}
\begin{IEEEproof}
Same with \textbf{Corollary~\ref{appro}}, but only LOS intra-cluster interferences exist.
\end{IEEEproof}
\end{corollary}

\begin{corollary}\label{corollaryd}
\emph{(Closed-form lower bound): Under \textbf{Assumption~\ref{assumption1}} and \textbf{Assumption~\ref{assumption2}}, \textbf{Theorem~2}} has a lower bound, which is shown as below.
\begin{align}\label{lower}
P_s \ge \sum\limits_{n = 1}^{{N_L}} {{{\left( { - 1} \right)}^{n + 1}} {N_L \choose n}} \frac{1}{{1 + \frac{{2\xi \psi \left( {\bar s - 1} \right)}}{{{\alpha _L}}}{{\left( {\frac{{\gamma_{th} {\eta _L}n}}{{{G_0}{N_L}}}} \right)}^{\frac{2}{{{\alpha _L}}}}}}},
\end{align}
\emph{where $\xi  = \sum\limits_{i = 1}^{N_G} {{b_i}} {\left( {{a_i}} \right)^{\frac{2}{{{\alpha _L}}}}}$, $\psi  = \int_0^\infty  {\left( {1 - \frac{1}{{{{\left( {y + 1} \right)}^{{N_L}}}}}} \right){y^{ - \frac{2}{{{\alpha _L}}} - 1}}dy}$.}
\begin{IEEEproof}
See Appendix~D.
\end{IEEEproof}
\end{corollary}
\begin{remark}\label{remark_5}
We observe that the coverage probability has a positive correlation with typical directivity gain $G_0$. On the contrary, it has the inverse correlation with the number of simultaneously transmitting devices $\bar{s}$ and modified expectation gain of antenna $\xi$. Significantly, the value of $\alpha _L$ should be larger than 2, otherwise, the $\psi$ will be infinite.
\end{remark}

\subsection{Closest Distribution Model}
Closest distribution model allows the typical user to access with the nearest transmitter in the typical cluster as we mentioned above. When we exploit the limited user¡¯s CSI which only contains the location information of devices, the nearest transmitter can be regarded as the energy enhancing device in the typical cluster as described in \emph{NNNF Selection Scheme} of NOMA networks~\cite{7445146}. In order to analyze the performance of this model, we will use Laplace transform to calculate the SINR coverage in the following part.
\subsubsection{Laplace Transform of Interference}
The same with uniform distribution model, we first derive the analytical expressions and approximations for intra-cluster interference with the Laplace transform.
\begin{lemma}\label{lemma6}
\emph{(Laplace transform of interference in the typical cluster): In this model, the typical link $r_1$ is the distance from the typical user to its nearest intra-transmitter. When $M\gg\bar{s}$, the $n$th conditional Laplace transform of interference in the typical cluster is given by}
\begin{align}\label{Lin1}
\begin{split}
&L_{I_{\mathop{intra}}}^n(s,r_1|{v_{c0}})={\exp\bigg({  { - (\bar s - 1)}}} \\
&\times {{{\int_{{r_1}}^\infty  {\left( {Q(s,{s_r}) + Z(s,{s_r})} \right)}{f_{{S_r}}}({s_r}|{v_{c0}},{r_1})d{s_r}} }\bigg)},
\end{split}
\end{align}
\begin{IEEEproof}
The proof process is same with \eqref{Lin}, but the distribution of simultaneous transmitters in the typical user is different. In this case, they follows \eqref{nearest} and the range is $[r_1,\infty)$.
\end{IEEEproof}
\end{lemma}
\subsubsection{Coverage Probability}
We use the same method as discussed in the uniform distribution model, the coverage probability is as follows
\begin{align}
\begin{split}
P_c=&{P_L}\left\{ {\frac{{{G_0}|{h_0}{|^2}{C_L}{r_1}^{ - {\alpha _L}}}}{{\left( {\sigma _n^2 + {I_{intra}} + {I_{inter}}} \right)}} > \gamma_{th} } \right\}\\
&+ {P_N}\left\{ {\frac{{{G_0}|{h_0}{|^2}{C_N}{r_1}^{ - {\alpha _N}}}}{{\left( {\sigma _n^2 + {I_{intra}} + {I_{inter}}} \right)}} > \gamma_{th} } \right\}.
\end{split}
\end{align}

\begin{theorem}
\emph{Using Laplace transform of intra/inter-cluster interference discussed above in \textbf{Lemma~\ref{lemma6}} and \textbf{Lemma~\ref{lemma4}}, we figure out the tight upper bound for the coverage probability in closest distribution model. }
\begin{align}\label{cloPc}
\begin{split}
P_c < &\int\limits_0^\infty  {\int\limits_0^\infty  {\left( {{X_1}\left( {{r_1},{v_{c0}}} \right) + {Y_1}\left( {{r_1},{v_{c0}}} \right)} \right)} }\\ &\times{f_{{R_1}}}({r_1}|{v_{c0}}){f_{{V_{c0}}}}({v_{c0}})d{r_1}d{v_{c0}},
\end{split}
\end{align}
\emph{where }
\begin{align}
\begin{split}
 &{X_1}\left( {{r_1},{v_{c0}}} \right)\\
 = &\sum\limits_{n = 1}^{{N_L}} {{{( - 1)}^{n + 1}}{ N_L \choose n }}\exp \left({-\varepsilon r_1}\right)\exp \left({ { - \frac{{\gamma_{th} {r_1}^{{\alpha _L}}{\eta _L}}}{{{C_L}{G_0}}}n\sigma _n^2} }\right)\\
&\times L_{{I_{intra}}}^n\left(\frac{{\gamma_{th} r_1^{{\alpha _L}}\eta_L}}{{{C_L G_0}}},{r_1}|{v_{c0}}\right)L_{{I_{inter}}}^n\left(\frac{{\gamma_{th} r_1^{{\alpha _L}}\eta_L}}{{{C_L G_0}}}\right),
\end{split}
\end{align}
\begin{align}
\begin{split}
&{Y_1}\left( {{r_1},{v_{c0}}} \right)=\sum\limits_{n = 1}^{{N_N}}{{( - 1)}^{n + 1}}\\
&\times {{ N_N \choose n }}(1-\exp \left({-\varepsilon r_1}\right))\exp \left({ { - \frac{{\gamma_{th} {r_1}^{{\alpha _N}}{\eta _N}}}{{{C_N}{G_0}}}n\sigma _n^2} }\right) \\
 &\times L_{{I_{intra}}}^n\left(\frac{{\gamma_{th} r_1^{{\alpha _N}}\eta_N}}{{{C_N G_0}}},{r_1}|{v_{c0}}\right)L_{{I_{inter}}}^n\left(\frac{{\gamma_{th} r_1^{{\alpha _N}}\eta_N}}{{{C_N G_0}}}\right).
 \end{split}
\end{align}
\begin{IEEEproof}
Same with \textbf{Theorem~\ref{theorem1}}, but the intra-cluster interference is different. The corresponding transmitter is the nearest one, so the simultaneous transmitters in the typical cluster is further than that in uniform distribution model. We use \eqref{Lin1} instead.
\end{IEEEproof}
\end{theorem}

\begin{corollary}\label{corollary6}
\emph{(Approximation of coverage probability): Under \textbf{Assumption~\ref{assumption1}} the tight upper bound for the probability of coverage is shown as below.}
\begin{align}
P_c^a<{\int\limits_0^\infty  {\left( {W({r_1}) + K({r_1})} \right)} } {f_{{R_1}}^a}(r_1)d{r_1},
\end{align}
\emph{where $f_{{R_1}}^a\left( {{r_1}} \right) = M \exp \left( {{ { - \frac{(M - 1){r_1^2}}{{4{\sigma ^2}}}} }}\right){f_D}\left( {{r_1}} \right)$.}
\begin{IEEEproof}
Same with \textbf{Corollary~\ref{appro}}, but the distance distribution of the typical link follows $f_{{R_1}}^a\left( {{r_1}} \right)$. The distribution $f_{{R_1}}^a\left( {{r_1}} \right)$ is proved by \textbf{Corollary~12} in~\cite{7446343}. We choose $k=1$ in this case.
\end{IEEEproof}
\end{corollary}
\begin{remark}\label{remark_6}
$f_{{R_1}}^a\left( {{r_1}} \right)$ is the PDF of distance between the typical user and the corresponding transmitter in closest distribution model, so with the same distance the probability of $r_1$ is higher than that of $r_0$ in uniform distribution model. In other words, the corresponding transmitter locates closer to the typical user in closest distribution model, which contributes to a higher coverage probability.
\end{remark}

\subsection{Closest LOS Model}
Instead of choosing the closest device in the previous model, closest LOS model is under the rule that the typical user communicates with the nearest LOS transmitter. If the complete CSI including the blockage information is available for the typical user, the received power can be enhanced by choosing the nearest D2D-Tx with an LOS link rather than NLOS link, as NLOS links experience higher path loss and severer channel fading than LOS links in mmWave networks~\cite{6932503}. Under this condition, closest LOS model is studied in this part. As discussed above, we also use Laplace transform to calculate the SINR coverage.
\subsubsection{Laplace Transform of Interference}
In this model, analytical expressions and approximations using the Laplace transform of intra-cluster interference are derived first.
\begin{lemma}\label{lemma7}
\emph{(Laplace transform of interference in the typical cluster): In the typical cluster the simultaneous transmitters are divided into two groups. One is LOS group and the other is NLOS group. In LOS group all transmitters are connecting with the typical user using LOS links and NLOS group is using NLOS links. Under the condition of that, the corresponding transmitter in this case is the nearest intra-transmitter in LOS group. When $M\gg\bar{s}$, the $n$th conditional Laplace transform of interference in the typical cluster is given by}
\begin{align}\label{LinL}
\begin{split}
&\dot{L}_{I_{\mathop{intra}}}^n(s,r_L|{v_{c0}})\\
=& \exp \bigg({  - ( {\bar s - 1} )\big(\int_{{r_L}}^\infty  {Q\left(s, {{s_L}} \right){f_{{S_L}}}({s_L}|{v_{c0}},{r_L})d{s_L}}}\\
&+ { \int_0^\infty  {Z\left( s,{{s_L}} \right)} {f_R}({s_L}|{v_{c0}})d{s_L}}\big)\bigg).
\end{split}
\end{align}
\begin{IEEEproof}
 The proof procedure is similar as \eqref{Lina}, but the distribution of simultaneous LOS transmitters in the typical cluster is different. In this model, they follow \eqref{nearestLOS} and the range is $[r_L,\infty)$.
\end{IEEEproof}
\end{lemma}

From \eqref{LinL}, we find that the LOS and NLOS group can be regarded as two independent non-homogeneous PCP. Moreover, LOS group will follow the same calculation process as discussed in closest distribution model and NLOS group will utilize the same method as in uniform distribution model.
\subsubsection{Coverage Probability}
As the corresponding transmitter connects to the typical user with an LOS link, there is no probability of typical NLOS link in this model. Utilizing the same method in other two scenarios.
\begin{align}
P_l=&{P_L}\left\{ {\frac{{{G_0}|{h_0}{|^2}{C_L}{r_L}^{ - {\alpha _L}}}}{{\left( {\sigma _n^2 + {I_{intra}} + {I_{inter}}} \right)}} > \gamma_{th} } \right\}.
\end{align}

\begin{theorem}
\emph{With the Laplace transform of intra-cluster interference \textbf{Lemma~\ref{lemma7}} and inter-cluster interference \textbf{Lemma~\ref{lemma4}}, we work out the tight upper bound for the coverage probability in closet LOS model as below.}
\begin{align}\label{LOSPc}
P_l< \int\limits_0^\infty  {\int\limits_0^\infty  {{X_L}\left( {{r_L},{v_{c0}}} \right)} } {f_{{R_L}}}({r_L}|{v_{c0}}){f_{{V_{c0}}}}({v_{c0}})d{r_L}d{v_{c0}},
\end{align}
\emph{where}
\begin{align}
\begin{split}
&{X_L}\left( {{r_L},{v_{c0}}} \right) = \sum\limits_{n = 1}^{{N_L}} {{{( - 1)}^{n + 1}}{ N_L \choose n }}\exp\left({ { - \frac{{\gamma_{th} {r_L}^{{\alpha _L}}{\eta _L}}}{{{C_L}{G_0}}}n\sigma _n^2} }\right)\\
 &\times \dot{L}_{{I_{{\rm{int}}ra}}}^n\left(\frac{{\gamma_{th} r_L^{{\alpha _L}}\eta_L}}{{{C_L G_0}}},{r_L}|{v_{c0}}\right)L_{{I_{inter}}}^n\left(\frac{{\gamma_{th} r_L^{{\alpha _L}}\eta_L}}{{{C_L G_0}}}\right).
\end{split}
\end{align}
\begin{IEEEproof}
 The proof procedure is similar as \textbf{Theorem~\ref{theorem1}}, but the corresponding transmitter is the nearest device with an LOS link. This part use \eqref{LinL} instead. In addition, we delete the $Y(r_0,v_{c0})$ in \eqref{Y} as there is no typical link with NLOS.
\end{IEEEproof}
\end{theorem}

\begin{corollary}\label{corollary7}
\emph{(Approximation of coverage probability): Under \textbf{Assumption~\ref{assumption1}} the tight upper bound for the probability of coverage is shown as below.}
\begin{align}
P_l^a<\int\limits_0^\infty  {W({r_L})} f_{{R_L}}^a({r_L})d{r_L},
\end{align}
\emph{where $f_{{R_L}}^a\left( {{r_L}} \right) = M{\left( {1 - \int_0^{{r_L}} {F_L^a(r)dr} } \right)^{M - 1}}F_L^a({r_L})$ and $F_L^a(r) = \exp \left( { - \varepsilon r} \right){f_D}\left( r \right)$.}
\end{corollary}

\subsection{Area Spectral Efficiency}
The ASE is the average bits transmitted per unit bandwidth per unit time and per unit area. We use Shannon's Capacity Formula when assuming that the D2D-Txs utilizes Gaussian Codebooks to calculate $ASE = \lambda {\log _2}\left( {1 + \gamma_{th} } \right){P_c}$, where $\lambda$ is the mean number of simultaneously active transmitters per unit area.

\begin{proposition}
\emph{The $ASE$ for three scenarios is same as below}
\begin{align}
ASE = \bar{s}\lambda_p {\log _2}\left( {1 + \gamma_{th} } \right){P},
\end{align}
\emph{where $P\in\{P_u, P_c, P_l\}$ is the coverage probability under three models shown as \eqref{uniPc}, \eqref{cloPc} and \eqref{LOSPc}, respectively.}
\end{proposition}

\begin{remark}\label{remark_7}
Note that there exists an optimum number of simultaneously transmitting devices, because more simultaneous transmitters potentially contribute to higher $ASE$, while it increases the interference essentially as well. As a result, $ASE$ is maximized by choosing optimum $\bar{s}$.
\begin{align}
ASE^{*} = opt(\bar{s})\lambda_p {\log _2}\left( {1 + \gamma_{th} } \right){P},
\end{align}
where $opt(\bar{s})$ is the optimum $\bar{s}$ that contributes to maximizing $ASE$ and $\bar{s}\in\{1,2,...,M\}$.
\end{remark}

\section{Numerical Results}\label{Numerical}

\subsection{General Network Simulation and Validation}
\begin{figure*}[t!]
\centering
\subfigure[Coverage probability versus various average number of simultaneous transmitters $\bar{s}$, with $\sigma=20$.]{\label{m} \includegraphics[width= 3.0in, height=2.25in]{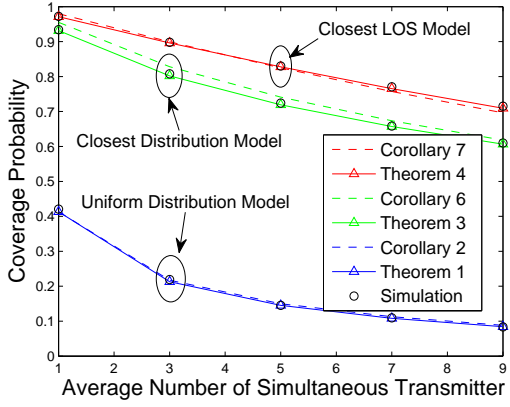}}
\subfigure[Coverage probability versus various pre-decided threshold of SINR $\gamma_{th}$, with the carrier frequency is 60 GHz, $\sigma=10$ and $\bar{s}=10$.]{\label{low_bound} \includegraphics[width= 3.0in, height=2.25in]{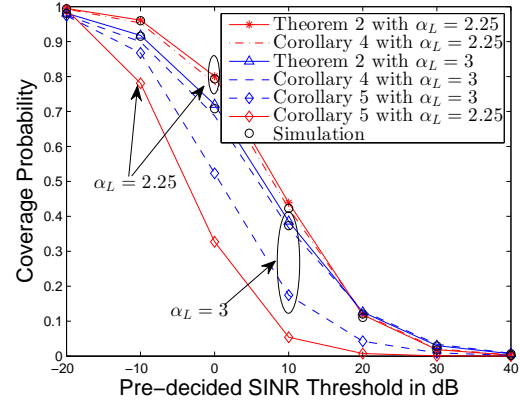}}
\caption{Simulation and validation}
\end{figure*}

We present the basic network settings in Table.~\ref{setting}~\cite{7446343,6932503}. In this paper, the reference distance is 1 meter and $C_L=C_N$. As shown in Fig.~\ref{m}, our analytical results match the simulations with negligible difference. Additionally, the \textbf{Corollary~\ref{appro},~\ref{corollary6}, and~\ref{corollary7}} are significantly tight to their corresponding simulation results. Closest LOS model performs the best among them and following is closest distribution model. Uniform distribution model is the worst regarding the coverage probability. This result is same with \textbf{Remark~\ref{remark_6}}.

In terms of the special case, Fig.~\ref{low_bound} illustrates that our closed-form lower bound has a reasonable distance apart from the simulation result, and it is capable to show the trends of our system with an efficient calculation process. Note that $\alpha_L$ should be more than 2, we employ carrier frequency at 60 GHz with $\alpha_L=2.25$~\cite{rappaport201238}. When comparing with $\alpha_L=3$, it shows that the lower bound will move closer to the simulation result with the increase of $\alpha_L$.
\begin{table}[h]
\centering
\caption{General Setting of Network ~\cite{7446343,6932503}}
\label{setting}
\begin{tabular}{|l|l|}
\hline
   Poisson cluster process region     & 1000 m $\times$ 1000 m\\ \hline
   Density of PPP     & $\lambda_P=150$ cluster/km$^2$\\ \hline
   Bandwidth per resource block    & $W=100$ MHz\\ \hline
   Path law for LOS links    & $\alpha_L=2$, $N_L=3$\\ \hline
   Path law for NLOS links     & $\alpha_N=4$, $N_N=2$\\ \hline
   Beam pattern for transmitters     & $G_{10dB, -10dB, 30^\circ}$\\ \hline
   Beam pattern for transmitters     & $G_{10dB, 0dB, 90^\circ}$\\ \hline
   Carrier frequency            & 28 GHz\\ \hline
   Average Distance of LOS      & 30 m\\ \hline
   Number of transmitters in one cluster & $M=40$\\ \hline
   Pre-decided SINR threshold            & $\gamma_{th}=20$ dB\\ \hline
\end{tabular}
\end{table}

\subsection{Impact of intra/inter-cluster interference}
\begin{figure*}[t!]
\centering
\subfigure[Coverage probability versus various average number of simultaneous transmitters $\bar{s}$, with $\gamma_{th}=10$ dB.]{\label{intra_inter} \includegraphics[width= 3.0in, height=2.25in]{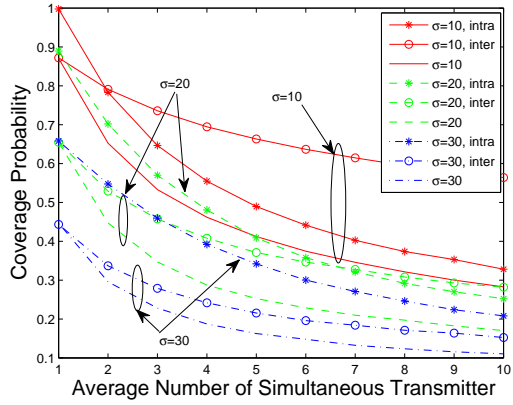}}
\subfigure[Coverage probability of the proposed i.i.d model and the practical scenario versus various average number of simultaneous transmitters $\bar{s}$, with $\gamma_{th}=10$ dB.]{\label{SIR_SINR} \includegraphics[width= 3.0in, height=2.25in]{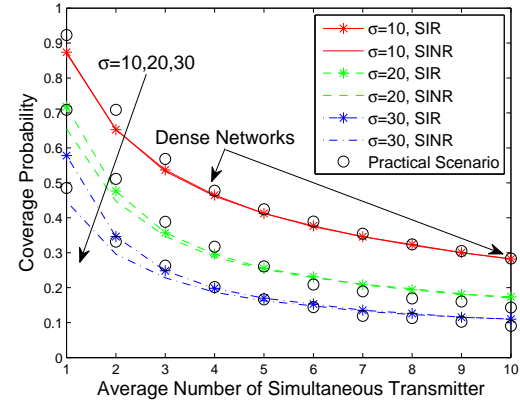}}
\caption{Structure property of clustered D2D mmWave networks}
\end{figure*}

We compare different coverage probabilities of 1) Only intra-cluster interference, 2) Only inter-cluster interference and 3) Both intra and inter-cluster interference with various scattering standard deviation $\sigma$. Obviously, when the average number of simultaneous transmitter $\bar{s}=1$, which means no interfering device in the typical cluster, the interferences are all from inter-clusters. Apart form $\bar{s}=1$, Fig.~\ref{intra_inter} demonstrates that when $\bar{s}$ increases, the intra-cluster interference will dominate our system. Therefore, an \emph{exchange number} of $\bar{s}$ exists to indicate whether the network is an intra-cluster interference limited or inter-cluster interference limited system. As shown in Fig.~\ref{intra_inter}, the exchange number is the cross of two kinds of interference lines. For instance, the exchange number for $\sigma=10$ is 2, which means when the $\bar{s}$ is less than 2, the network is an inter-cluster interference limited system, while $\bar{s}$ is more than 2, it will be an intra-cluster interference limited system. With the augment of $\sigma$, the exchange number will be larger.

\subsection{Impact of Noise and Blockage Model}
As we use thermal noise in the model, it is interested in analyzing whether the noise plays a critical role in our system. Fig.~\ref{SIR_SINR} illustrates that when $\sigma=10$, noise is negligible so that proposed network is an interference-limited system. With the increase of $\sigma$, the noise will never be discarded at the small $\bar{s}$ region. It turns to be a noise-effective system. As the D2D has the content centric nature, which means $\sigma$ is small in practice, this clustered D2D mmWave communication network can be regarded as an interference-limited system.

When the density of the devices in a cluster becomes high, one obstacle will block all transmitters behind it in reality, but our i.i.d LOS assumption with the stochastic blockage model is still accurate. We assume a practical scenario with a non-i.i.d blockage process. All obstacles are located at the edge of an LOS ball with radius $R_s$, where $p(R_s)=0.5$. It means that transmitters located within the ball are the LOS nodes, while that outside the ball will transmit information to the typical user with an NLOS link~\cite{6932503}. Fig.3(b) shows that the difference between our assumed random blockage model and the practical scenario is fairly close. especially in the dense networks area with $\sigma=10$ and high $\bar{s}$, thereby validating our analysis.

\subsection{Impact of LOS interference and Average Distance of LOS}
In this part, we focus on the LOS interference and average distance of LOS. As shown in Fig.~\ref{LOSdis}, three models are mainly affected by LOS interference, which means NLOS interference is negligible in our system. Average distance of LOS is also an important variable for mmWave network, it reflects the density of the blockage in the area. Fig.~\ref{LOSdis} illustrates that when the average distance of LOS raises, the difference between closest distribution model and closest LOS model will be gradually eliminated.

\subsection{Impact of Beamforming}
\begin{figure*}[t!]
\centering
\subfigure[Coverage probability versus various average number of simultaneous transmitters $\bar{s}$, with $\sigma=20$ and $\gamma_{th}=20$ dB.]{\label{LOSdis} \includegraphics[width= 3.0in, height=2.25in]{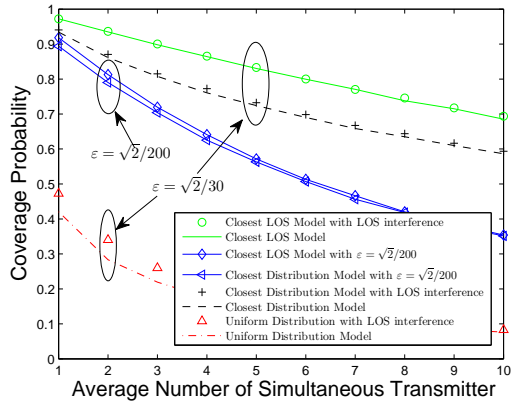}}
\subfigure[Coverage probability versus various pre-decided threshold of SINR $\gamma_{th}$, with $\bar{s}=3$ and $\sigma=10$.]{\label{different_antenna} \includegraphics[width= 3.0in, height=2.25in]{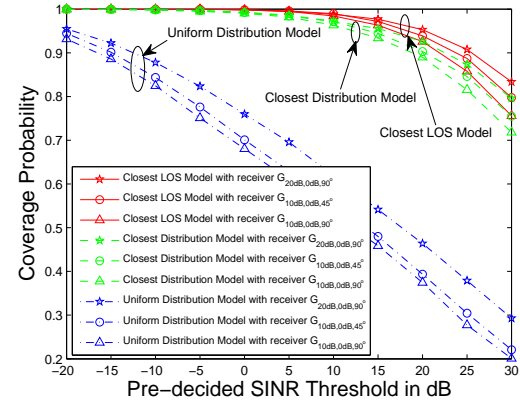}}
\caption{Antenna beamforming and blockage effects of clustered D2D mmWave networks}
\end{figure*}

We assume the antenna parameter of the corresponding transmitter is fixed as $G_{10dB, -10dB, 30^\circ}$. Comparing the coverage probability with different antenna patterns of the typical user, Fig.~\ref{different_antenna} shows that when the side lobe gain is stationary, the coverage probability arises with the increase of main lobe gain or decrease of main lobe beamwidth, because the large main lobe gain contributes to the large received power at the typical user, and small main lobe beamwidth reduces the probability of large interference $b_1$ in Table.~\ref{tab1}. Fig.~\ref{different_antenna} also demonstrate that although three scenarios take the noise, NLOS links and inter-cluster interference into account, they have the same trend with the special case in uniform distribution model as mentioned in \textbf{Remark~\ref{remark_5}}. In a word, the coverage probability has a positive correlation with the typical directivity gain $G_0$.

\subsection{Performance of ASE}
\begin{figure*}[t!]
\centering
\subfigure[Coverage probability versus various average number of simultaneous transmitters $\bar{s}$ for three different models.]{\label{ASE1} \includegraphics[width= 3.0in, height=2.25in]{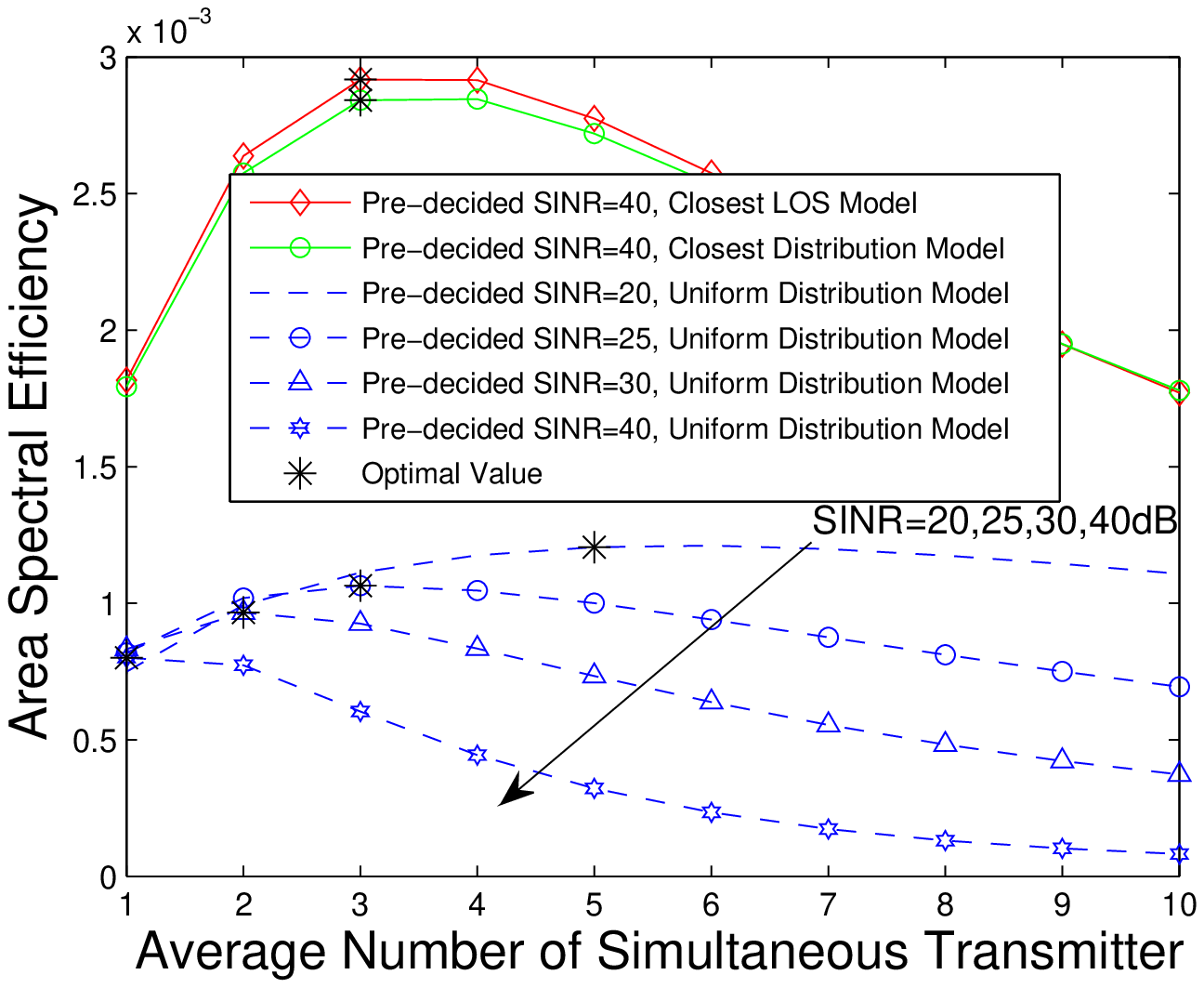}}
\subfigure[Coverage probability versus various pre-decided threshold of SINR $\gamma_{th}$, with $\sigma=30$.]{\label{different_f} \includegraphics[width= 3.0in, height=2.25in]{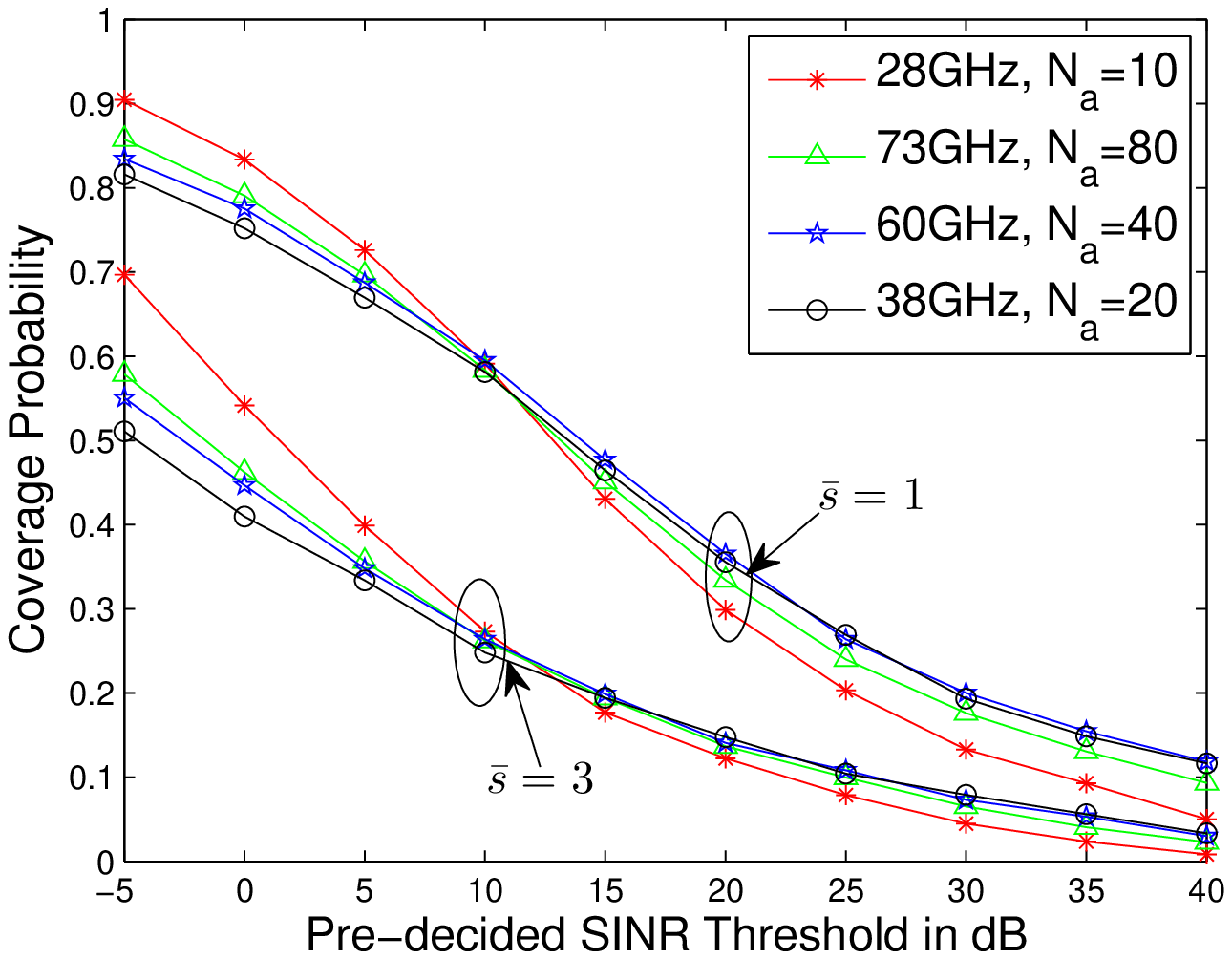}}
\caption{The performance of ASE and different carrier frequencies in clustered D2D mmWave networks}
\end{figure*}

We present ASE with the average number of simultaneous transmitters $\bar{s}$ in this part. In Fig.~\ref{ASE1}, it shows the ASE performance of three different scenarios. The optimal number of $\bar{s}$ can be easily worked out from Fig.~\ref{ASE1}, because they are convex functions and the highest point is the optimal number as we discussed in \textbf{Remark~\ref{remark_7}}. When the pre-decided threshold of SINR increases, the optimal number decreases. Moreover, it is obvious that closest LOS model and closest distribution model have larger ASEs than uniform distribution model.

\subsection{Performance of Different Carrier Frequencies}
We concentrate on carrier frequencies at 28 GHz, 38 GHz, 60 GHz, and 73GHz. Based on the practical channel measurements, the LOS and NLOS path loss exponents are shown in Table~\ref{exponent}~\footnote{The number of antenna elements here is an estimation just for illustrating the different performance of various carrier frequencies.}~\cite{deng201528,rappaport201238}. Although the number of antenna arrays is different across the various carrier frequencies, it has limited impact on the SINR coverage probability. The reason is that the effect of antenna scales is canceled when considering the interferences since both received power and interferences will be simultaneously enhanced at the same level. Therefore large antenna scales can only compensate the loss by noise in our system.The performance of different carrier frequencies is shown in Fig. 5(b). Based on the discussion in part C of Section V, both in the noise-effective system ($\bar{s}=1$) and interference-limited system ($\bar{s}=3$), 28 GHz is the best for under $\gamma_{th}=10 dB$ SINR regions, while 38 GHz outperforms others for high SINR regions.
\begin{table}[h]
\centering
\caption{Path Loss Exponent for mmWave Outdoor Channels~\cite{deng201528,rappaport201238}}
\label{exponent}
\begin{tabular}{|c|c|c|c|c|}
\hline
   Path Loss Exponent    & 28G    &38G    &60G     &73G\\ \hline
   LOS $\alpha_L$     & 2    &2   &2.25     &2  \\ \hline
   Strongest NLOS $\alpha_N$     & 3    &3.71    &3.76     &3.4   \\ \hline
   Number of antenna elements $N_a$  &10 &20 &40 &80 \\ \hline
\end{tabular}
\end{table}

\section{Conclusion}\label{conclusion}
In this paper, the performance of clustered device-to-device mmWave communication is examined. The stochastic geometry is utilized to model three different scenarios. Specifically, closest LOS model performs the best in terms of coverage probability. Although closest LOS model has a higher coverage probability than closest distribution model, the deviation between these two scenarios is gradually eliminated when the average distance of LOS increases. We analytically demonstrate that the coverage probability has the inverse correlation the number of simultaneously transmitting devices, but it has a positive correlation with the typical directivity gain.  As discussed in the previous sections, our frame is an interference-limited system and NLOS interference is negligible due to content centric nature. Maximum ASE can be achieved by choosing the optimal number of simultaneous transmitters. After comparing different carrier frequencies, we conclude that 28 GHz is the best choice for low SINR region and 38 GHz is the best for high SINR region. As the optimal number of simultaneous transmitting devices in terms of ASE may not be the best value for required coverage probability, there is a trade-off between these two parameters. We will study this optimization in our future work.

\numberwithin{equation}{section}
\section*{Appendix~A: Proof of Lemma~\ref{lemma3}} \label{Appendix:A}
\renewcommand{\theequation}{A.\arabic{equation}}
\setcounter{equation}{0}

The Laplace transform of the intra-cluster interference is
\begin{align}\label{A1}
\begin{split}
L_{{I_{intra}}}^n(s|{v_{c0}}) ={\mathbb{E}}\left[ {\exp\left({ { - snI_{intra}^L} }\right)} \right]{\mathbb{E}}\left[ {\exp\left({ { - snI_{intra}^N} }\right)} \right],
\end{split}
\end{align}
where $I_{intra}^L$ and $I_{intra}^N$ are the intra-cluster interference from LOS and NLOS links respectively. The proof of ${\mathbb{E}}\left[ {\exp\left({ - snI_{intra}^L}\right)} \right]$ is
\begin{align}\label{A2}
& {\mathbb{E}}\left[ {\exp\left( { { - snI_{intra}^L} }\right)} \right]\nonumber\\
\mathop  = \limits^{(a)}& {\mathbb{E}_{{G_l}}}\left[ {{\mathbb{E}_{x_d}}}{\left[ { {\frac{{{N_L}^{{N_L}}}}{{{{\left( {sn{G_l}{C_L}||{x_{c0}} + {x_d}|{|^{ - {\alpha _L}}} + {N_L}} \right)}^{{N_L}}}}}} } \right]} \right]\nonumber\\
\mathop  = \limits^{(b)}& {\mathbb{E}_{{G_l}}}\Bigg[ {\sum\limits_{j = 0}^{M - 1} {{{\bigg( {\int_{{\mathbb{R}^2}} {\frac{{{N_L}^{{N_L}}}}{{{{\left( {sn{G_l}{C_L}||{x_{c0}} + {x_d}|{|^{ - {\alpha _L}}} + {N_L}} \right)}^{{N_L}}}}}}} }}}}\nonumber\\
&\times {{{{{{{f_{{X_d}}}\left( {{x_d}} \right)d{x_d}} } \bigg)}^j}\frac{{{\raise0.7ex\hbox{${{{\left( {\bar s - 1} \right)}^j} e^{{  { - \left( {\bar s - 1} \right)} }}}$} \!\mathord{\left/
 {\vphantom {{{{\left( {\bar s - 1} \right)}^j}e^{ { - \left( {\bar s - 1} \right)} }} {j!}}}\right.\kern-\nulldelimiterspace}
\!\lower0.7ex\hbox{${j!}$}}}}{{\sum\limits_{k = 0}^{M - 1} {{\raise0.7ex\hbox{${{{\left( {\bar s - 1} \right)}^k} e^{{ { - \left( {\bar s - 1} \right)} }}}$} \!\mathord{\left/
 {\vphantom {{{{\left( {\bar s - 1} \right)}^k}e^{  { - \left( {\bar s - 1} \right)} }} {k!}}}\right.\kern-\nulldelimiterspace}
\!\lower0.7ex\hbox{${k!}$}}} }}} } \Bigg]\nonumber\\
\mathop  = \limits^{(c)} &\exp\left({  { - \left( {\bar s - 1} \right) {\int_0^\infty  {Q\left(s, r \right){f_R}\left( {r|{v_{c0}}} \right)dr} } } }\right),
\end{align}
where (a) is computing the moment generating function of Gamma random variable $|h_l|^2$; (b) follows the fact that the locations of active intra-cluster transmitters are independent; (c) is the expectation of antenna gain under the assumption $\bar{s}\ll M$. Using the same process, we are capable to work out that:
\begin{align}\label{A3}
\begin{split}
&{\mathbb{E}}\left[ {\exp\left({  - snI_{intra}^N}\right)} \right]\\
=&\exp\left({  { - \left( {\bar s - 1} \right) {\int_0^\infty  {Z\left(s, r \right){f_R}\left( {r|{v_{c0}}} \right)dr} } } }\right).
\end{split}
\end{align}

Then by substituting \eqref{A2} and \eqref{A3} into \eqref{A1}, we obtain \eqref{Lin}. The proof is complete.

\section*{Appendix~B: Proof of Lemma~\ref{lemma4}} \label{Appendix:B}
\renewcommand{\theequation}{B.\arabic{equation}}
\setcounter{equation}{0}

The Laplace transform of the inter-cluster interference is
\begin{align}\label{B1}
&L_{{I_{inter}}}^n(s)= {\mathbb{E}}\left[ {\exp\left( { { - snI_{inter}^L} }\right)} \right]{\mathbb{E}}\left[ {\exp\left( { { - snI_{inter}^N} }\right)} \right],
\end{align}
where $I_{inter}^L$ and $I_{inter}^N$ are the inter-cluster interference from LOS and NLOS links respectively. The proof of ${\mathbb{E}}\left[ {\exp\left( { - snI_{inter}^L}\right)} \right]$ is
\begin{align}\label{B2}
\begin{split}
 &\mathbb{E}\left[ {\exp \left( { - snI_{inter}^L} \right)} \right]\\
 \mathop  = \limits^{(a)}& {\mathbb{E}_{\Phi _P^L}}\left[ { {\exp\left({ { - \bar s {\int_0^\infty  {Q\left(s, w \right){f_W}\left( {w|{v}} \right)dw} } } }\right)} } \right]\\
\mathop  = \limits^{(b)}& {\exp\left({ - 2\pi {\lambda _p}\int_0^\infty  {\left( {1 - {{\mathop{\rm e}\nolimits} ^{ - \bar s\int_0^\infty  {Q\left( {s,w} \right){f_W}\left( {w|v} \right)dw} }}} \right)vdv} }\right)},
\end{split}
\end{align}
where (a) is following the same method in \emph{Appendix~A}, but the number of interfering transmitters in intra-cluster is $\bar{s}$ in this case; (b) follows the probability generating functional of PPP~\cite{stoyanstochastic,liu2016nonorthogonal}. Then it changes the coordinates to polar. Using the same process, we are able to figure out that
\begin{align}\label{B3}
\begin{split}
&\mathbb{E}\left[ {\exp\left({ { - snI_{inter}^N} }\right)} \right]\\
=&\exp\left({  { - 2\pi {\lambda _p}\int_0^\infty  {\left( {1 - e^ { { - \bar s {\int_0^\infty  {Z\left(s, w \right){f_W}\left( {w|v} \right)dw} } } }} \right)vdv} } }\right).
\end{split}
\end{align}

Then by substituting \eqref{B2} and \eqref{B3} into \eqref{B1}, we obtain \eqref{Lout}. The proof is complete.

\section*{Appendix~C: Proof of Theorem~\ref{theorem1}} \label{Appendix:C}
\renewcommand{\theequation}{C.\arabic{equation}}
\setcounter{equation}{0}

When the typical user is associated with a transmitter with an LOS link, the interference is composed of simultaneously transmitting devices from both the typical cluster $I_{intra}$ and other clusters $I_{inter}$. The LOS conditional probability of coverage is shown below:
\begin{align}\label{C1}
\begin{split}
                                &{P_L}\left\{ {\frac{{{G_0}|{h_0}{|^2}{C_L}{r_0}^{ - {\alpha _L}}}}{{\left( {\sigma _n^2 + {I_{intra}} + {I_{inter}}} \right)}} > \gamma_{th} }\right\}\\
                                 =&{P_L}\left\{ {|{h_0}{|^2} > \frac{{\gamma_{th} {r_0}^{{\alpha _L}}\left( {\sigma _n^2 + {I_{intra}} + {I_{inter}}} \right)}}{{{C_L}{G_0}}}}| r_0\in \mathbb{R}_L \right\} \\
 \mathop  < \limits^{(a)} & 1 - \mathbb{E}\left[ {{{\left( {1 - e^ { { - \frac{{\gamma_{th} {r_0}^{{\alpha _L}}{\eta _L}}}{{C_L{G_0}}}\left( {{{\sigma _n^2}+I_{intra}} + {I_{inter}}} \right)} }} \right)}^{{N_L}}}}| r_0\in \mathbb{R}_L \right]\\
 \mathop {\rm{ = }}\limits^{(b)}& \sum\limits_{n = 1}^{{N_L}} {{{( - 1)}^{n + 1}}{N_L\choose n}}e^{{-\varepsilon r_0}} e^{ { { - \frac{{\gamma_{th} {r_0}^{{\alpha _L}}{\eta _L}}}{{{C_L}{G_0}}}n\sigma _n^2} }}\\
 &\times\mathbb{E}\left[ {e^{  { - \frac{{\gamma_{th} {r_0}^{{\alpha _L}}{\eta _L}}}{{C_L{G_0}}}n{I_{intra}}} } }\right] \mathbb{E}\left[ {e^{  { - \frac{{\gamma_{th} r_0^{{\alpha _L}}{\eta _L}}}{{C_L{G_0}}}n{I_{inter}}} } }\right]\\
 \mathop  = \limits^{(c)}       & \int\limits_0^\infty  {\int\limits_0^\infty  {X({r_0},{v_{c0}})} } {f_R}({r_0}|{v_{c0}}){f_{{V_{c0}}}}({v_{c0}})d{r_0}d{v_{c0}},
\end{split}
\end{align}
where (a) is a tight upper bound when $N_L$ is small~\cite{alzer1997some}, that is ${P_L}\left\{ {|{h_0}{|^2} < \psi } \right\} < {\left( {1 - e^ { { - \psi {\eta _L}} }} \right)^{{N_L}}}$; (b) is from Binomial theorem when $N_L$ is an integer and $\mathbb{R}_L$ is the set of $r_0$ which belongs to the LOS group; (c) is from the results of \emph{Appendix~A} and \emph{Appendix~B}. The NLOS conditional probability of coverage can be calculated in the same way and the expression is
\begin{align}\label{C2}
\begin{split}
&{P_N}\left\{ {\frac{{{G_0}|{h_0}{|^2}C_N{r_0}^{ - {\alpha _N}}}}{{\left( {{\sigma _n^2 + I_{intra}} + {I_{inter}}} \right)}} > \gamma_{th} } \right\}\\
< & \int\limits_0^\infty  {\int\limits_0^\infty  {Y({r_0},{v_{c0}})} } {f_R}({r_0}|{v_{c0}}){f_{{V_{c0}}}}({v_{c0}})d{r_0}d{v_{c0}}.
\end{split}
\end{align}

Then by substituting \eqref{C1} and \eqref{C2} into \eqref{P}, we obtain \eqref{uniPc}. The proof is complete.

\section*{Appendix~D: Proof of Corollary~\ref{corollaryd}} \label{Appendix:D}
\renewcommand{\theequation}{D.\arabic{equation}}
\setcounter{equation}{0}

Under \textbf{Assumption~\ref{assumption1}}, the tight upper bound for coverage probability of special case is given by
\begin{align}\label{D1}
 P_s< &\int\limits_0^\infty  \sum\limits_{n = 1}^{{N_L}} {{{( - 1)}^{n + 1}}{ N_L \choose n }} \ddot{L}_{{I_{intra}}}^n\left(\frac{{\gamma_{th} r_0^{{\alpha _L}}\eta_L}}{{{C_L G_0}}}\right) f_{{R}}({r_0})d{r_0}.
\end{align}

The Laplace transform of the intra-cluster interference in this special case is as follows
\begin{align}\label{D2}
\begin{split}
  &\ddot{L}_{I_{intra}}^n(s)\\
  =&\int_{{\mathbb{R}^2}} {\exp \bigg( - \left( {\bar s - 1} \right)} \sum\limits_{i = 1}^{N_G} {{b_i}}  \\
   &\times \int_{{\mathbb{R}^2}} {\big(1 - \frac{{{N_L}^{{N_L}}}}{{{{\left( {{a_i}sn||{x_{c0}} + {x_d}|{|^{ - {\alpha _L}}} + {N_L}} \right)}^{{N_L}}}}}\big)} {f_{{X_d}}}\left( {{x_d}} \right)d{x_d}\bigg)  \\
   & \times {f_{{X_d}}}\left( {{x_{c0}}} \right)d{x_{c0}}\\
  \mathop  \ge \limits^{(a)} &{\exp\left({ - \frac{{\xi \psi \left( {\bar s - 1} \right)}}{{2{\sigma ^2}{\alpha _L}}}{{\left( {\frac{{\gamma_{th} {\eta _L}n}}{{{G_0}{N_L}}}} \right)}^{\frac{2}{{{\alpha _L}}}}}r_0^2}\right)},
\end{split}
\end{align}
where (a) follows Jensen's inequality and Young's inequality (see \emph{Appendix~D} in~\cite{7446343}). And $\xi  = \sum\limits_{i = 1}^{N_G} {{b_i}} {\left( {{a_i}} \right)^{\frac{2}{{{\alpha _L}}}}}$, $\psi  = \int_0^\infty  {\left( {1 - \frac{1}{{{{\left( {y + 1} \right)}^{{N_L}}}}}} \right){y^{ - \frac{2}{{{\alpha _L}}} - 1}}dy}$. $\psi$ is a constant when $\alpha _L>2$.

Then by substituting \eqref{D2} into \eqref{D1}, we obtain \eqref{lower}. The proof is complete.

\bibliographystyle{IEEEtran}
\bibliography{mybib}

\end{document}